\documentclass[12pt,draftcls,onecolumn]{IEEEtran}
\usepackage{amssymb,amsmath,amsopn,amsfonts,graphicx,colordvi,CJK,epstopdf,subfigure}
\usepackage[justification=centering]{caption}
\usepackage{amsmath,amssymb,amsthm,mathrsfs,amsfonts,amstext,empheq}
\usepackage{graphicx,subfigure}
\usepackage[comma,numbers,square,sort&compress]{natbib}
\usepackage{textcomp}
\usepackage{tikz}
\usetikzlibrary{shapes,arrows,shadows}
\usepackage{verbatim}
\usepackage{amsmath,bm,times}
\input epsf

\tikzset{%
  block/.style    = {draw, thick, rectangle, minimum height = 2.5em,
    minimum width = 3em, fill=red!20, drop shadow},
  sum/.style      = {draw, circle, node distance = 2cm, fill=blue!20}, % Adder
  input/.style    = {coordinate}, % Input
  output/.style   = {coordinate} % Output
}
\tikzstyle{sensor}=[draw, fill=blue!20, text width=4em,
    text centered, minimum height=2.5em,drop shadow]
\tikzstyle{ann} = [above, text width=5em, text centered]
\tikzstyle{wa} = [block, thick, rectangle, fill=red!20, minimum height = 3em,
    minimum width = 3em, drop shadow]
\tikzstyle{sc} = [sensor, text width=13em, fill=red!20,
    minimum height=10em, rounded corners, drop shadow]
\tikzstyle{nt}=[sensor, text width=6.5em, fill=red!20,
    minimum height=4.5em, rounded corners, drop shadow]
\def\baselinestretch{1.4}

\def\a{\alpha}
\def\b{\beta}

\def\ba{\begin{array}}
\def\ea{\end{array}}
\def\ban{\begin{eqnarray*}}
\def\ean{\end{eqnarray*}}
\def\bd{\begin{description}}
\def\ed{\end{description}}
\def\be{\begin{equation}}
\def\ee{\end{equation}}
\def\bna{\begin{eqnarray}}
\def\ena{\end{eqnarray}}

\def\d{\delta}

\newtheorem{definition}{Definition}
\newtheorem{remark}{Remark}
\newtheorem{theorem}{Theorem}
\newtheorem{lemma}{Lemma}

\begin{document}
\title{Distributed Averaging With Random Network Graphs and Noises}

\author{Tao Li,~\IEEEmembership{Senior Member,~IEEE}, and
Jiexiang Wang

\IEEEcompsocitemizethanks{\IEEEcompsocthanksitem
 This work was supported by the National Science Found for Outstanding Young Scholars, National Natural Science Foundation of China under Grant 61522310 and the Shanghai Rising-Star Program under grant 15QA1402000. Please address all the correspondences to Tao Li: Phone: +86-21-54342646-318, Fax: +86-21-54342609, Email: tli@math.ecnu.edu.cn.
\IEEEcompsocthanksitem Tao Li is with Department of Mathematics, East China Normal University, Shanghai 200241, China.
%He was with School of Mechatronic Engineering and Automation, Shanghai University, Shanghai 200072, China.
\IEEEcompsocthanksitem Jiexiang Wang is with School of Mechatronic Engineering and Automation, Shanghai University, Shanghai 200072, China.
}}

\maketitle

\begin{abstract}
We consider discrete-time distributed averaging algorithms over  multi-agent networks with measurement
noises and time-varying random graph flows. Each agent updates its state by relative states between neighbours with both additive and multiplicative measurement noises. The network structure is modeled by time-varying random digraphs, which may be spatially and temporally dependent. By developing difference inequalities of proper stochastic Lyapunov function,  the algebraic graph theory and martingale convergence theory, we obtain sufficient conditions for stochastic approximation type algorithms to achieve mean square and almost sure average consensus. We prove that all states of agents converge to a common variable in mean square and almost surely if the graph flow is conditionally balanced and uniformly conditionally jointly connected. The mathematical expectation of the common variable is right the average of initial values, and the upper bound of the mean square steady-state error is given quantitatively related to the weights, the algorithm gain and the energy level of the noises.
\end{abstract}


\section{Introduction}
In real networked systems, there exist various kinds of uncertain factors, such as channel noises, channel fading, random link failures and recreations.
In recent years,  stochastic multi-agent networks have attracted great attention from scholars in various fields
and become an active interdisciplinary research subject.
For stochastic multi-agent networks, the distributed averaging is one of the most fundamental problems and has wide application background, such as distributed computation (\cite{IEEEhowto:lynch}-\cite{IEEEhowto:xiao1}), distributed filtering
(\cite{IEEEhowto:Olfati}-\cite{IEEEhowto:OlfatiShamma}), information fusion over wireless sensor networks (\cite{IEEEhowto:xiao2}), distributed learning and optimization (\cite{Sayed2014}-\cite{Tsitsiklis1986}) and load balancing (\cite{AFJV2015TIT}) etc.

The measurement/communication noises affect not only the decision-making of each individual agent, but also the overall performance of the whole system. %This brings essential difficulties for each agent's own decision and the integral analysis of the closed-loop system.
Generally, measurement/communication noises are divided into two categories: additive and multiplicative noises. The additive noise corrupts the signal in the form of superposition regardless of the signal's own intensity. Differently, the multiplicative noise is coupled with the signal together. For example, the effects of coherent fading can be modeled by multiplicative noises in imaging radar systems (\cite{058}). For distributed averaging with additive measurement noises, Huang and Manton (\cite{041}) proposed the discrete-time stochastic approximation type average-consensus protocol, and gave sufficient conditions for mean square consensus under fixed undirected graphs. Li and Zhang (\cite{016}) studied  the continuous-time distributed averaging algorithm with additive measurement noises and obtained necessary and sufficient conditions for mean square average-consensus under fixed balanced digraphs.
For distributed averaging with multiplicative measurement noises, Li $et~al.$ (\cite{021}) considered the average consensus under fixed undirected graphs with nonlinear noise intensity functions, and gave the necessary and sufficient conditions for mean square average consensus.
Ni and Li (\cite{020}) considered distributed consensus with multiplicative measurement noises where the noise intensities are the absolute values of relative states.

Besides measurement/communication noises, the network structure of a multi-agent network often randomly changes due to  packet dropouts, link/node failures and recreations, which are particularly serious for wireless networks. The random switching of the network structure has a strong impact on the convergence and performance of distributed averaging algorithms. This topic also attracts extensive attentions from the community of distributed averaging.
Distributed averaging and consensus with independent identically distributed (i.i.d.) graph flows were considered in \cite{047}-\cite{061}. Especially, Bajovi\'{c} $et~al.$ \cite{BXMS2013TSP} proved that the exact convergence rate in probability is exponentially fast for products of i.i.d. symmetric stochastic matrices.  The cases with ergodic stationary and finite state homogeneous Markov chain type graph flows were considered in (\cite{049}) and (\cite{045}), respectively, which both obtained necessary and sufficient conditions for almost sure consensus.
Liu $et~al.$ \cite{026} and Touri and  Nedic \cite{052} studied distributed consensus with more general random graph flows.
Liu $et~al.$ \cite{026} obtained sufficient conditions for $L_p$ consensus assuming that the $\delta-$graph contains a spanning tree.
Touri and  Nedic \cite{052}  gave a more general condition for the convergence of weak periodic random matrix sequences.

Most of the above literature considered the effect of random changing of network structures or measurement noises on distributed algorithms separately. In real networks, various kinds of uncertainties may co-exist, for example, there may be additive measurement noises and channel fading accompanied with random link changes. Many scholars have long been committed to developing distributed averaging algorithms with comprehensive uncertainties, establishing convergence conditions and quantitative relations between algorithm performances and network parameters. However,
as far as we know, few results were obtained on distributed averaging with the above three kinds of random uncertainties integrated together.
 Li and Zhang \cite{015} considered the distributed averaging with additive measurement/communication noises and deterministic switching graph flows. They established the necessary and sufficient condition for mean square average consensus under fixed digraphs and the jointly-containing-spanning-tree condition for mean square and almost sure average consensus under switching digraphs.  Rajagopal and Wainwright \cite{Ram} studied the distributed averaging with additive storage noises, additive communication noises and data-constrained communication.
Kar and Moura \cite{062} gave sufficient conditions for almost sure consensus under Markov chain type graph flows with the fixed mean graph and additive measurement noises.  Huang $et~al.$ \cite{013} considered the case with spatial-temporal-independent additive  measurement noises and random link gains under Markov and deterministic switching network graph flows. They obtained sufficient conditions for mean square and almost sure consensus.
Aysal and  Barner \cite{0250} proposed a model of general consensus dynamics and gave conditions for almost sure convergence under additive disturbances and randomly switching graphs.  Patterson $et~al.$ \cite{PBA2010TAC} considered distributed averaging with spatial-temporal-independent random link failures and random input noises. They  gave the exponential mean square convergence rate for mean square average-consensus assuming that the underlying mean graph is always undirected and connected.
Wang and Elia \cite{0240} focused on the system fragilities caused by communication constraints (additive input noises, communication delay and fading channels). They established the tight relation among  uncertainties of network channels, robust mean square stability and the appearance of Le\'{v}y flight, and gave conditions for mean square weak consensus without additive input noises. Furthermore,
Wang and Elia \cite{023} studied how the model parameters affect the appearance of complex behaviour and  provided an expression to verify the system stability. Long $et~al.$ \cite{022} considered distributed consensus with multiplicative noises and randomly switching graphs assuming that the mean graph is fixed and connected.

In this paper, we consider discrete-time multi-agent distributed averaging algorithms with both additive and multiplicative measurement noises under time-varying random graph flows. %The information networks among agents are modeled by randomly time-varying digraph processes.
The time-varying algorithm gain is adopted to attenuate the noises. By constructing difference inequalities of proper stochastic Lyapunov function and tools of algebraic graph theory and martingale convergence theory,  we obtain sufficient conditions for the distributed approximation type algorithm to achieve mean square and almost sure average consensus. In detail, we prove that all states of agents converge to a common random variable in mean square and almost surely if the random graph flow is \emph{conditionally balanced and uniformly conditionally jointly connected}. The mathematical expectation of the variable is right the average of initial states of agents. Moreover, we give an upper bound of the mean square steady-state error quantitatively related to the edge weights, the algorithm gain, the number of agents, the agents' initial states, the second-order moment and the intensity coefficients of the noises. %The simulation verifies our results.
Compared with the relevant literature, main contributions of our paper are summarized as follows.

I. The measurement model covers both cases with additive and multiplicative noises.
%In \cite{020},\cite{021} and \cite{022},  the case with only multiplicative noises were considered, so a fixed algorithm gain can ensure exponential convergence, which is the key to obtain the almost convergence of the algorithm.
On one  hand, different from the case with only multiplicative noises, due to the introduction of the time-varying algorithm gain to attenuate the additive noises, the closed-loop system becomes a time-varying stochastic system, and the exponential convergence of the expectation of the Lyapunov energy function, which is essential to obtain the almost sure consensus conditions in \cite{021}-\cite{020} and \cite{022}, can not be used.
%On the basis of \cite{015}, we further develop the stochastic Lyapunov method and get the almost sure average consensus by the nonnegative supermartingale convergence theorem and the properties of the algorithm gain.
On the other hand, different from the case with only additive measurement noises (\cite{041}-\cite{016}, \cite{015}),  multiplicative noises relying on the relative states between agents make states and noises coupled together in a distributed information structure. This leads to an additional martingale term with coupled states and network graphs in the system centroid equation. The estimation for the term leads to more complex
analysis for the closed-loop steady-state error.
To these ends, we construct  difference inequalities of proper stochastic Lyapunov function. Firstly, by martingale convergence theory, we prove the boundedness of the closed-loop states. Then  by substituting the boundedness back into the difference inequality, we obtain mean square average consensus. Furthermore, by tools of martingale convergence theory, we obtain almost sure average consensus.
It is worth pointing out that though Wang and Elia (\cite{0240}-\cite{023}) considered both additive input noises and Bernoulli fading channels, they used fixed algorithm gain and only obtained mean square weak consensus in absence of the additive input noises.  %Thus, our work can not be contained by it.
In addition, different from the most existing literature, the noises in this paper are allowed to be spatially and temporally dependent.

II. In this paper, the stochastic Lyapunov flow method is further developed for the case with time-varying random graph flows. Li and Zhang \cite{015} considered deterministic switching digraphs and proved that if the network graph flow switches among instantaneously balanced digraphs and is jointly-connected over fixed length consecutive time intervals, then mean square and almost sure consensus are achieved. In Huang \cite{Huang2012TAC}, the lengths of the time intervals, over which the network is jointly connected, can randomly vary but must be bounded with probability one. In fact,
The network graph flow conditions given in \cite{015} and \cite{Huang2012TAC} are essentially deterministic type conditions. However, for random graph flows, it is very difficult to verify whether their sample paths satisfy such kinds of conditions with probability one.  Particularly, the sample paths of Markovian switching graphs do not satisfy those conditions. In this paper, the network structure among agents is modeled by more general random graph flows. The {\em generalized weighted adjacency matrices} are not required to have special statistical properties, such as independency with identical distribution, Markovian switching or stationarity, etc.  By introducing the concept of conditional digraph and martingale convergence theory, we establish the \emph{uniformly conditionally jointly connected condition} to ensure stochastic average consensus. The jointly-connected conditions with respect to i.i.d. graph flows, Markovian and deterministic switching graph flows in the existing literature are all special cases of our condition.   Moreover, different from \cite{015}, which assumed that the digraphs are balanced, we only require that the conditional digraph is balanced; and different from \cite{062} and \cite{022}, we do not require fixed mean graph.

III.  In real networks, there exist not only cooperative, but also antagonistic relations between agents (\cite{055}-\cite{057}). Such relations can be modeled by links with positive or negative weights, respectively.   In most existing literature on distributed averaging, nonnegative edge weights are required.  Liu $et~al$  \cite{026} and Touri and
Nedic \cite{052} studied noise-free consensus algorithms under random graph flows, and required nonnegative edge weights.  Porfiri and Stilwell (\cite{047}) considered noise-free distributed consensus with arbitrary weights in a sampled-data setting, however, the network graph flow is required to be i.i.d. and the mean graph is always connected. In this paper, we show that under the uniformly conditionally jointly connected condition, even though the random edge weights take negative values at some time instants, mean square and almost sure consensus can also be achieved.
%So  our results  can not be covered by \cite{026} and \cite{052} even for the noise-free case.

The remaining parts of this paper are arranged as follows. Section II gives preliminaries and the problem formulation. Section III gives main results.  In Section IV, for two special cases of Markovian switching graph flows with countable states and independently switching graph flows with uncountable states, the sufficient conditions for mean square and almost average consensus are given.
%Section V presents the simulation examples to demonstrate the theoretical results.
Section V gives the concluding remarks and some future topics.

Notation and symbols:\\
$\textbf{1}_N$: $N$-dimensional vector with all ones;\\
$\textbf{0}_N$, $N$-dimensional vector with all zeros; \\
$I_N$: $N$-dimensional identity matrix;\\
%$J_N$:  $N$-dimensional square matrix with $\frac{1}{N}$ entries;\\
$O_{m\times n}$:   $m\times n$  dimensional zero matrix; \\
$\mathbb R$: set of real numbers;\\
$A\ge B$: matrix $A-B$ is positive semi-definite;\\
$A\succeq B$: matrix $A-B$ is a nonnegative matrix;\\
$A^T$: transpose of matrix $A$;\\
$diag(B_1,...,B_n)$: block diagonal matrix with entries being $B_1$,...,$B_n$;\\
$\|A\|$:  2-norm of matrix $A$;\\
$\|A\|_{F}$:  Frobenius-norm of matrix $A$;\\
$E(\xi)$: mathematical expectation of random variable $\xi$;\\
$Var(\xi)$: variance of $\xi$;\\
$|S|$: the cardinal number of set $S$;\\
$\lceil x\rceil$: the minimal integer greater than or equal to real number  $x$;\\
$\lfloor x \rfloor$: the maximal integer smaller  than or equal to  $x$;\\
$b_n=O(r_n)$: $\limsup_{n\to\infty}\frac{|b_n|}{r_n}<\infty$, where $\{b_n$, $n\ge0\}$ is a real sequence and $\{r_n$, $n\ge0\}$ is a positive real sequence;\\
$b_n=o(r_n)$: $\lim_{n\to\infty}\frac{b_n}{r_n}=0$;\\
$\mathcal F_\eta(k)=\sigma(\eta(j)$, $0\le j\le k)$, $k\ge0$, $\mathcal F_\eta(-1)=\{\Omega$, $\emptyset\}$, where $\{\eta(k)$, $k\ge0\}$ is a random vector or matrix sequence.

\section{Preliminaries and Problem Formulation}
\subsection{Preliminaries}
Let the triple $\mathcal{G}=\{\mathcal{V},\mathcal{E}_{\mathcal{G}},\mathcal{A}_{\mathcal{G}}\}$
be a weighted digraph, where
$\mathcal{V}=\{1,...,N\}$ is the node set with node $i$ representing agent $i$; $\mathcal{E}_{\mathcal{G}}$ is the edge set, and $(j,i)\in\mathcal{E}_\mathcal G$ if and only if agent $j$ can send information to agent $i$ directly. Denote the neighbourhood of agent $i$ by $N_i=\{j\in \mathcal{V}|(j,i)\in
\mathcal{E}_{\mathcal{G}}\}$.
We call $\mathcal{A}_{\mathcal{G}}=[a_{ij}]\in \mathbb{R}^{N\times N}$ \emph{the generalized weighted adjacency matrix of $\mathcal{G}$ }, where $a_{ij}\not=0 \Leftrightarrow j\in N_i$. Since $\mathcal{E}_\mathcal G$ is uniquely determined by $\mathcal{A}_{\mathcal{G}}$, the digraph can also be denoted by the pair $\mathcal{G}=\{\mathcal{V},\mathcal{A}_{\mathcal{G}}\}$.
The in-degree and out-degree of agent $i$ are denoted by $deg_{in}(i)=\sum_{j=1}^Na_{ij}$ and $deg_{out}(i)=\sum_{j=1}^Na_{ji}$, respectively.
We call $L_{\mathcal{G}}=\mathcal{D}_{\mathcal{G}}-\mathcal{A}_{\mathcal{G}}$
 \emph{the generalized  Laplacian matrix of $\mathcal{G}$ }, where $\mathcal{D}_{\mathcal{G}}=diag(deg_{in}(1),...,deg_{in}(N))$. By the definition, $L_{\mathcal G}\textbf{1}_N=\textbf{0}_N$.
If $deg_{in}(i)=deg_{out}(i)$, $\forall~i\in\mathcal{V}$, then $\mathcal{G}$ is balanced.
We call $\widetilde{\mathcal{ G}}=\{\mathcal{V},{\mathcal{E}_{\widetilde{\mathcal{G}}}},{\mathcal{A}_{\widetilde{\mathcal{G}}}}\}$ the reversed digraph of $\mathcal{G}$, where $(i,j)\in{\mathcal{E}_{\widetilde{\mathcal{G}}}}$ if and only if $(j,i)\in\mathcal{E}_{\mathcal{G}}$ and ${\mathcal{A}_{\widetilde{\mathcal{G}}}}=\mathcal{A}_{\mathcal{G}}^T$. Then $\widehat{\mathcal G}=\{\mathcal V,\mathcal{E}_{\mathcal{G}}\cup{\mathcal{E}_{\widetilde{\mathcal{G}}}},\frac{\mathcal A_{\mathcal G}+\mathcal A_{\mathcal G}^T}{2}\}$ is called the symmetrized graph of $\mathcal G$.
Denote $\hat L_\mathcal G=\frac{L_\mathcal G+L^T_\mathcal G}{2}$. If $a_{ij}\ge0 $, $\forall~i$, $j\in\mathcal{V}$, then $\mathcal{A}_{\mathcal{G}}$ and $L_{\mathcal G}$ defined before degenerate to the  weighted adjacency matrix and Laplacian matrix in common sense, and $\hat L_\mathcal G$ is the Laplacian matrix of $\widehat{\mathcal G}$ if and only if $\mathcal G$ is balanced (\cite{cpnas2004}).

The union digraph of
$\mathcal{G}_{1}=\{\mathcal{V},\mathcal{E}_{\mathcal{G}_{1}},\mathcal{A}_{\mathcal{G}_{1}}\}$ and
$\mathcal{G}_{2}=\{\mathcal{V},\mathcal{E}_{\mathcal{G}_{2}},\mathcal{A}_{\mathcal{G}_{2}}\}$ with the common node set $\mathcal{V}$
is denoted by $\mathcal{G}_{1}+\mathcal{G}_{2}=\{\mathcal{V},
\mathcal{E}_{\mathcal{G}_{1}}\cup\mathcal{E}_{\mathcal{G}_{2}},
\mathcal{A}_{\mathcal{G}_{1}}+\mathcal{A}_{\mathcal{G}_{2}}\}$.
By  the definition of $L_\mathcal G$, we know that
$L_{\sum_{j=1}^{k}\mathcal{G}_{j}}=\sum_{j=1}^{k}L_{\mathcal{G}_{j}}$.
A sequence of edges $(i_{1}, i_{2})$, $(i_{2}, i_{3})$, ..., $(i_{k-1},i_{k})$ is called a directed path from $i_{1}$
to $i_k$. If for all $i$, $j\in\mathcal{V}$, there exists a directed path from $i$ to $j$, then
$\mathcal{G}$ is strongly connected.

\subsection{Problem Formulation}
Consider a multi-agent system of $N$ agents whose information structure is described by a
time-varying random digraph flow $\{{\mathcal G}(k)=\{\mathcal{V},\mathcal{E}_{\mathcal{G}(k)},\mathcal{A}_{\mathcal{G}(k)}\},k\ge0\}$.
We consider the following distributed averaging algorithm:
\bna
\label{discretemodel}
x_{i}(k+1)=x_i(k)+c(k)\sum_{j\in N_i(k)}a_{ij}(k)(y_{ji}(k)-x_i(k)),\  k\ge0,\ i\in\mathcal V,
\ena
where $x_{i}(k)\in\mathbb R$ is the state of agent $i$ at time instant $k$, and the initial states
$x_i(0)$, $i=1,2,..., N$ are deterministic variables. Here, $N_i(k)$ denotes the neighbourhood of agent $i$ at time instant $k$, $c(k)$ is the time-varying algorithm gain, and
$y_{ji}(k)$ denotes the measurement of agent $j$'s state by its neighbouring node $i$ at time instant $k$, which is given by
\bna \label{meaequa}
y_{ji}(k)=x_j(k)+f_{ji}(x_j(k)-x_i(k))\xi_{ji}(k),\ i\in \mathcal V,\ j\in N_i(k).
\ena
%$y_{ji}(k)$可以看作第 $i$ 个自主体对第 $j$ 个Agent的状态 $x_j(k)$ 所作的量测，
where $\{\xi_{ji}(k), k\geq0\}$ represents the measurement noise sequence on channel $(j,i)$ and $f_{ji}(x_j(k)-x_i(k))$ is the noise intensity function.
The combination of systems (\ref{discretemodel}) and (\ref{meaequa}) is called the distributed stochastic approximation type consensus algorithm (\cite{041}, \cite{015}, \cite{062}). Let $\xi(k)=[\xi_{11}(k),...,\xi_{N1}(k);...;\xi_{1N}(k),...,\xi_{NN}(k)]^T$, where $\xi_{ji}(k)\equiv0$ if $j\notin N_{i}(k)$ all for $k\geq0$. For the measurement model (\ref{meaequa}) and the algorithm gain $c(k)$, we have the
following assumptions.%make several assumptions for the measurement model (\ref{meaequa}) and time-varying algorithm gain as follows：

\vskip 0.2cm

\textbf{(A1)} The noise intensity function $f_{ji}(\cdot)$ is a mapping from  $\mathbb R$ to $\mathbb R$. There exist positive constants $\sigma_{ji}$ and $b_{ji}$, $i$, $j\in\mathcal{V}$, such that  $\left|f_{ji}(x)\right|\le\sigma_{ji}|x|+b_{ji}$, $\forall$ $x\in\mathbb R$.

\vskip 0.2cm

\textbf{(A2)} The noise process $\{\xi(k),\mathcal F_\xi(k)$, $k\ge0\}$ is a sequence of vector-valued martingale differences and there exists a positive constant $\b$ such that $\sup_{k\ge0}E\left[\|\xi(k)\|^2|\mathcal F_\xi(k-1)\right]\le\beta$~ a.s.%$\mathcal F_\xi(k)=\sigma(\xi(j), 0\le j\le k)$.

\vskip 0.2cm

\textbf{(A3)} $c(k)>0$, $\forall~k\ge0$, $\sum_{k=0}^\infty c(k)=\infty$, $\sum_{k=0}^\infty c^2(k)<\infty$.

\vskip 0.2cm

~\textbf{(A4)} $c(k)\downarrow0$, $c(k)=O(c(k+h))$, $k\to\infty$, $\forall~ h>0$.

\vskip 0.2cm

\begin{remark}
{\rm Assumption \textbf{(A1)} shows that the measurement model (\ref{meaequa}) covers both cases of additive  and multiplicative measurement noises. Here, $b_{ji}$, $i$, $j\in\mathcal{V}$  and $\sigma_{ji}$, $i$, $j\in\mathcal{V}$  are additive and multiplicative noise intensity coefficients, respectively. This measurement model is suitable for many practical multi-agent systems. For example, in the formation control, to achieve the final desired formation, agents need to get the position information of neighbours. This information acquisition process is usually corrupted by additive noises, such as electromagnetic interference. In addition, the larger the distance between agents is, the more unreliable the received information is, which can be modeled as multiplicative noises in the form of $\sigma_{ji}|x_j(k)-x_i(k)|\xi_{ji}(k)$.

The measurement models with additive noises in \cite{041}-\cite{016}, \cite{015} and those with multiplicative noises in \cite{021}-\cite{020} and \cite{022} are both special cases of model (\ref{meaequa}). In detail, the measurement model in  \cite{015} is
$y_{ji}(k)=x_j(k)+\xi_{ji}(k),j\in N_i(k)$.
The measurement model in \cite{021} is
$y_{ji}(k)=x_j(k)+f_{ji}(x_j(k)-x_i(k))\xi_{ji}(k),j\in N_i(k)$,
where $|f_{ji}(x_j(k)-x_i(k))|\le\sigma_{ji}|x_j(k)-x_i(k)|$.
The measurement model in \cite{020} and \cite{022} is
$y_{ji}(k)=x_j(k)+\sigma_{ji}|x_j(k)-x_i(k)|\xi_{ji}(k),j\in N_i(k)$.
Obviously, the noise intensity functions of the three models above all satisfy  \textbf{(A1)}.}
\end{remark}

%\vskip 0.2cm
%
%\begin{remark}
%{\rm In this paper, we assume that the overall noises constitute a martingale difference sequence without the requirement that the noises are spatial-temporal independent as in the most existing literatures (\cite{020}-\cite{021}, \cite{0240}-\cite{023}).
%%However, this weaker assumption leads to new difficulties in analyzing the closed-loop system, where the coupled term of states and noises can not be simply separated as the case with independent noises.  We take a full advantage of properties of the conditional expectation and work it out efficiently.
%}
%\end{remark}

\vskip 0.2cm

\begin{remark}
{\rm If only multiplicative measurement noises are considered, existing literature showed that the fixed algorithm gain can ensure strong consensus (\cite{021}-\cite{020}), however, such  fixed-gain algorithm can not ensure strong consensus if additive and multiplicative measurement noises co-exist. Here, we adopt the decaying algorithm gain $c(k)$ to attenuate the additive noises. In the field of distributed algorithms,  Assumption \textbf{(A3)} ensures that  $c(k)$ vanishes with a proper rate for attenuating noises and meanwhile the algorithm does not converge too early.}
\end{remark}

\vskip 0.2cm

Let $X(k)=[x_1(k),\cdots,x_{N}(k)]^{T}$,
%$\xi(k)=[\xi_1(k),\cdots,\xi_N(k)]^T$ where $\xi_i(k)=[\xi_{i1}(k),\cdots\\\xi_{iN}(k)]$,
$D(k)=diag(\alpha_1^T(k),...,\alpha_N^T(k))$ with $\alpha_i^T(k)$ being the $i $th row of  $\mathcal A_{\mathcal G(k)}$,
${Y}(k)=diag\left(f_1(k),\cdots,f_N(k)\right)$, where $f_i(k)=diag(f_{1i}(x_1(k)-x_i(k))$, $\cdots$, $f_{Ni}(x_N(k)-x_i(k)))$.
Substituting (\ref{meaequa}) into (\ref{discretemodel}) leads to the closed-loop system in the compact form
\bna\label{driven2}
X(k+1)=(I_N-c(k)L_{\mathcal G(k)})X(k)+c(k)D(k)Y(k)\xi(k).
\ena\par

\begin{definition}(\cite{015})
\label{dstrongconsensusaverage}
For systems (\ref{discretemodel}) and  (\ref{meaequa}), if for any given $X(0)
\in\mathbb{R}^{N}$, there exists a random variable $x^{*}$, such that $E(x^{*})=\frac{1}{N}\sum_{j=1}^{N}x_j(0)$, $Var(x^{*})<\infty$,
$\lim_{k\rightarrow\infty}E[x_{i}(k)-x^{*}]^2=0$,  $i\in\mathcal V$, and
$\lim_{k\to\infty}x_i(k)=x^*$  a.s., $i\in\mathcal V$,
then we say that systems (\ref{discretemodel}) and  (\ref{meaequa}) achieve mean square and  almost sure average consensus.
\end{definition}

In this paper, we aim at giving the conditions under which systems (\ref{discretemodel}) and (\ref{meaequa}) can achieve mean square and almost sure average consensus based on the models formulated above, i.e., the random digraph flow and the measurement model with both additive and multiplicative noises. The following section gives the main results of this paper.

\section{main results}

We first introduce the concept of \emph{conditional digraphs}. We call
$E[\mathcal A_{\mathcal G(k)}|\mathcal F_{\mathcal A}(m)]$, $m\leq k-1$, the  \emph{conditional generalized weighted adjacency matrix} of $\mathcal A_{\mathcal G(k)}$ with respect to $\mathcal F_{\mathcal A}(m)$, and call its associated random graph the \emph{conditional digraph} of $\mathcal G(k)$ with respect to $\mathcal F_{\mathcal A}(m)$, denoted by $\mathcal G(k|m)$, i.e.,  $\mathcal G(k|m)=\{\mathcal V, E[\mathcal A_{\mathcal G(k)}|\mathcal F_{\mathcal A}(m)]\}.$

In this section, we consider the random graph flow with balanced conditional digraphs as follows:
$$\Gamma_1=\Big\{\{\mathcal G(k),k\ge0\}|E[\mathcal A_{\mathcal G(k)}|\mathcal F_{\mathcal A}(k-1)]\succeq O_{N\times N}~\mbox{a.s.}, ~\mathcal G(k|k-1)~\mbox{is balanced a.s.}, ~ k\ge0\Big\}.$$

We have the following assumption on the random graph flow and the measurement noises.

~\textbf{(A5)} The random graph flow $\{{\mathcal G(k)},k\ge0\}$ and the noise process $\{\xi(k),k\ge0\}$ are mutually independent.

Let $J_N=\frac{1}{N}\textbf{1}\textbf{1}^T$ and $P_N=I_N-J_N$. Denote the consensus error vector $\d(k)=P_NX(k)$ and the Lyapunov energy function $V(k)=\|\d(k)\|^2$. For any given $k\ge0$ and positive integer $h$, denote
\bna\label{uniformalgebriacconnectivityaddadd1}
\lambda_{k}^h=\lambda_2\left(\sum_{i=k}^{k+h-1}E[ \hat{L}_{\mathcal G(i)}|\mathcal F_{\mathcal A}(k-1)]\right),
\ena
where $\lambda_2(\cdot)$ denotes the second smallest eigenvalue. Since $E[ \hat{L}_{\mathcal G(i)}|\mathcal F_{\mathcal A}(k-1)]$ is a real symmetric matrix a.s., $\lambda_{k}^h$ is well defined.

We are now in the position for the main result.

\begin{theorem}\label{mainresults}
For systems (\ref{discretemodel})-(\ref{meaequa}) and the associated random graph flow $\{\mathcal G(k),k\ge0\}\in\Gamma_1$, assume that

(a) Assumptions \textbf{(A1)}-\textbf{(A5)} hold;

(b) there exist deterministic positive integer $h$ and positive constants  $\theta$ and $\rho_0$, such that

~~~~(b.1) $\inf_{m\ge0}{{\lambda}}_{mh}^h\ge\theta$\ a.s.,

~~~~(b.2) $\sup_{k\ge0}\left[E[\|L_{\mathcal G(k)}\|^{2^{max\{h,2\}}}|\mathcal F_{\mathcal A}(k-1)]\right]^{\frac{1}{2^{max\{h,2\}}}}\le \rho_{0}$\ a.s.\\
Then as $k\to\infty$,  the consensus error $\delta(k)$ vanishes in mean square and almost surely, i.e., $\lim_{k\to\infty}E[V(k)]=0$ and
$\lim_{k\to\infty}V(k)=0$, a.s. Moreover,
all states $x_i(k)$, $i\in\mathcal V$,  converge to a common random variable $x ^*$, in mean square and almost surely, with
$E(x^*)=\frac{1}{N}\sum_{j=1}^{N}x_j(0)$ and
\bna\label{013}
Var(x^*)\le\frac{4c\b b^2\rho_{1}}{N^2}+\frac{8\widetilde{c} \b\sigma^2\rho_{1}}{N^2}+\frac{2c\rho_2 q_x}{N},
\ena
where
\ban
&&c=\sum_{k=0}^\infty c^2(k),\widetilde{c}=\sum_{k=0}^\infty E[V(k)] c^2(k),\sigma=\max_{1\le i,j\le N}\{\sigma_{ji}\},b=\max_{1\le i,j\le N}\{b_{ji}\},\cr
&&
%\overline q_x=\max\{ q_x,\|X(0)\|^2\},
q_x=\exp\left\{c\rho_{0}^2\right\}\Big(\|X(0)\|^2+2c\b\rho_{1}(2\sigma^2 q_v+b^2)\Big),\cr
&&q_v=\exp\left\{c(\rho_{0}^2 +4\rho_{1}\beta\sigma^2)\right\}\Big(V(0)+2c\b\rho_{1} b^2\Big),\cr%&&q_v=\exp(\rho(1+8\beta\sigma^2)\sum_{i=0}^\infty c^2(i))\{V(0)+4\rho b^2\b
&&\mbox{$\rho_{1}~$ and $\rho_2$ are constants satisfying}\cr
%\overline q_v=\max\{ q_v,V(0)\},
&&\sup_{k\ge0}E\left[|\mathcal E_{\mathcal G(k)}|\max_{1\le i,j\le N}a_{ij}^2(k)|\mathcal F_{\mathcal A}(k-1)\right]\leq\rho_{1},\ a.s.\cr
&&\max_{1\le i\le N}\sup_{k\ge0}E\left[\left(\sum_{j=1}^Na_{ij}(k)-\sum_{j=1}^Na_{ji}(k)\right)^2\Big|\mathcal F_{\mathcal A}(k-1)\right]\le\rho_2,\ a.s.
%&&\mbox{$d^*$  satisfying}: \sup_{k\ge0}\max_{1\le i\le N} N_i(k)\le d^*,a.s.\cr%&&\mbox{$\tau$  satisfying}: \sup_{k\ge0}E[\|D^T(k)D(k)\||\mathcal F_{\mathcal A}(k-1)]\le\tau, a.s.\cr
%&&\mbox{$q_1,q_2$满足}:\sup_{k\ge0}\max_{1\le i,j\le N}E(a_{ij}(k)|\mathcal F_{\mathcal A}(k-1))\le q_1,\sup_{k\ge0}\max_{1\le i,j\le N}E(a_{ij}^2(k)|\mathcal F_{\mathcal A}(k-1))\le q_2,a.s.\cr
\ean
\end{theorem}

\noindent
\textbf{Proof:} Firstly, if condition (b.2) holds, noting that
\ban
|\mathcal E_{\mathcal G(k)}|\max_{1\le i,j\le N}a_{ij}^2(k)\leq N(N-1)\max_{1\le i,j\le N}a_{ij}^2(k)\leq N(N-1)\|\|L_{\mathcal G(k)}\|_{F}^2,
\ean
by the equivalence of 2-norm and Frobenius norm of matrices and the conditional Lyapunov inequality, we know that the deterministic constants $\rho_{1}$ and
$\rho_{2}$ are both well defined.

This theorem is proved by 6 \textbf{Steps} as follows.

\noindent
\textbf{Step~1}: To prove $\sup_{k\geq0}E[V(k)]<\infty$.

By (\ref{driven2}) and the definition of $\d(k)$, we have
\ban\label{random1}
\delta(k+1)&=&P_N(I_N-c(k)L_{\mathcal G(k)})X(k)+c(k)P_ND(k)Y(k)\xi(k)\cr
&=&\d(k)-c(k)P_N L_{\mathcal G(k)}X(k)+c(k)P_N D(k)Y(k)\xi(k).
\ean
 By  the  definition of $L_{\mathcal G(k)}$, it follows that $L_{\mathcal G(k)}J_N=\textbf{0}_N$,  and so $L_{\mathcal G(k)}X(k)=L_{\mathcal G(k)}\d(k)$. Then from the above, we have
\bna\label{random111}
\delta(k+1)=(I_N-c(k)P_N L_{\mathcal G(k)})\delta(k)+c(k)P_ND(k)Y(k)\xi(k),
\ena
which together with the   definition of $V(k)$ leads to
\bna\label{dgks}
V(k+1)
&\le& V(k)-2c(k)\d^T(k)\frac{ L_{\mathcal G(k)}^TP_N^T+P_N L_{\mathcal G(k)}}{2}\d(k)+c^2(k)\|L_{\mathcal G(k)}\|^2\|\d(k)\|^2\cr
&&+~c^2(k)\xi^T(k)Y^T(k)D^T(k)P_ND(k)Y(k)\xi(k)\cr
&&+~2c(k)\xi^T(k)Y^T(k)D^T(k)P_N(I_N-c(k)P_NL_{\mathcal G(k)})\d(k).
\ena
By Lemma \ref{independence} and Assumption \textbf{(A2)}, we know that
\bna\label{dadfk}
E[\xi^T(k)Y^T(k)D^T(k)P_N(I_N-c(k)P_NL_{\mathcal G(k)})\d(k)]=0.
\ena
Noting that $\mathcal G(k|k-1)$ is balanced a.s., by Assumption \textbf{(A5)}, %$k\ge0$
we get
\ban
E\Big[\frac{ L_{\mathcal G(k)}^TP_N^T+P_N L_{\mathcal G(k)}}{2}\Big|\mathcal F_{\xi,\mathcal A}(k-1)\Big]&=&E\Big[\frac{ L_{\mathcal G(k)}^TP_N^T+P_N L_{\mathcal G(k)}}{2}\Big|\mathcal F_{\mathcal A}(k-1)\Big]\cr
&=&E[\hat L_{\mathcal G(k)}|\mathcal F_{\mathcal A}(k-1)]\ge O_{N\times N} ~a.s.,
\ean
and then by $\delta(k)\in\mathcal F_{\xi,\mathcal A}(k-1)$, we have
\bna\label{dkkfnal}
E\left[\d^T(k)\frac{ L_{\mathcal G(k)}^TP_N^T+P_N L_{\mathcal G(k)}}{2}\d(k)\right]\ge 0.
\ena
By Assumption \textbf{(A1)} and the definitions of $Y(k)$ and $V(k)$, we get
\begin{eqnarray}
\label{noiseintensity}
\|Y(k)\|^2%&=&\lambda_{\max}[Y^2(k)]\cr
&=&\max_{1\le i,j\le N}\left(f_{ji}(x_j(k)-x_i(k))\right)^2\cr
&\leq&\max_{1\le i,j\le N}[2\sigma^2(x_j(k)-x_i(k))^2+2b^2]\cr
&\le&4\sigma^2\max_{1\le j,i\le N}\Bigg[ \Big(x_j(k)-\frac{\sum_{i=1}^Nx_i(k)}{N}
\Big)^2+\Big(x_i(k)-\frac{\sum_{i=1}^Nx_i(k)}{N}\Big)^2\Bigg]+2b^2\nonumber\\
&\le&4\sigma^2\sum_{j=1}^N\Big(x_j(k)-\frac{\sum_{i=1}^Nx_i(k)}{N}\Big)^2+2b^2\cr
&=&4\sigma^2V(k)+2b^2.
\end{eqnarray}
%By the  direct calculation and the conditions in the theorem,
%\bna
%&&E[\|D^T(k)D(k)\||\mathcal F_{\mathcal A}(k-1)]\cr
%&&=E[\max_{1\le i\le N}\sum_{j=1}^Na^2_{ij}(k)|\mathcal F_{\mathcal A}(k-1)]=\sum_{j=1}^N
%E[\max_{1\le i\le N}a^2_{ij}(k)|\mathcal F_{\mathcal A}(k-1)]\le N\rho_{1}, a.s.
%\ena
This together  with  Assumptions \textbf{(A2)}, \textbf{(A5)} and Lemma \ref{independence} gives
\bna\label{dafmalka}
&&E[\xi^T(k)Y^T(k)D^T(k)P_ND(k)Y(k)\xi(k)]\cr&\le& E[\|Y(k)\|^2\|\xi(k)\|^2\|D^T(k)D(k)\|]\cr
&=&E\{E[\|Y(k)\|^2\|\xi(k)\|^2\|D^T(k)D(k)\||\mathcal F_{\mathcal A,\xi}(k-1)]\}\cr
&=&E\{\|Y(k)\|^2E[\|\xi(k)\|^2|\mathcal F_{\mathcal A,\xi}(k-1)]E[\|D^T(k)D(k)\||\mathcal F_{\mathcal A}(k-1)]\}\cr
&\le& \b E\{(4\sigma^2V(k)+2b^2)E[\|D^T(k)D(k)\||\mathcal F_{\mathcal A}(k-1)]\}\cr
&=&\b E\{(4\sigma^2V(k)+2b^2)E[\lambda_{max}(D^T(k)D(k))|\mathcal F_{\mathcal A}(k-1)]\}\cr
&=&\b E\left\{(4\sigma^2V(k)+2b^2)E\bigg[\max_{1\leq i\leq N}\lambda_{max}(\alpha_{i}(k)\alpha_{i}^T(k))|\mathcal F_{\mathcal A}(k-1)\bigg]\right\}\cr
&=&\b E\left\{(4\sigma^2V(k)+2b^2)E\bigg[\max_{1\leq i\leq N}\textbf{tr}(\alpha_{i}^T(k)\alpha_{i}(k))|\mathcal F_{\mathcal A}(k-1)\bigg]\right\}\cr
&\leq&\b E\left\{(4\sigma^2V(k)+2b^2)E\bigg[|\mathcal E_{\mathcal G(k)}|\max_{1\le i,j\le N}a_{ij}^2(k)|\mathcal F_{\mathcal A}(k-1)\bigg]\right\}\cr
&\le& 4\sigma^2\b\rho_{1} E[V(k)]+2b^2\b \rho_{1}.
\ena
From the above, taking the mathematical expectation on both sides of (\ref{dgks}), by (\ref{dadfk}), (\ref{dkkfnal})  and  condition (b.2), we get
\bna \label{energy28}
E[V(k+1)]
&\le&[ 1+c^2(k)(\rho_{0}^2+4\b\sigma^2 \rho_{1})]E[V(k)]
+2b^2\b \rho_{1} c^2(k),\ k\ge0.
\ena
This together with Assumption \textbf{(A3)} and Lemma \ref{robbins} gives that $E[V(k)]$ is bounded (regarding $E[V(k)]$ as $x(k)$ in Lemma \ref{robbins}).

\noindent
\textbf{Step~2}: To prove $E[V(k)]\to0$, $k\to\infty$.

Let
$\Phi(m,n)=(I_N-c(m-1)P_NL_{\mathcal G(m-1)})\cdots(I_N-c(n)P_NL_{\mathcal G(n)}),m\ge n,\Phi(n,n)=I_N$.
From (\ref{random111}) and by iterative calculations, we get
\ban\label{qqq}
\delta((m+1)h)=\Phi((m+1)h,mh)\delta(mh)+\tilde\xi_{m}^{mh},\ m\ge0,
\ean
where
\bna
\label{tildexi}
\tilde\xi_{m}^{mh}=\sum_{j=mh}^{(m+1)h-1}c(j)\Phi((m+1)h,j+1)P_N D(j)Y(j)\xi(j).
\ena
By the  definition of $V(k)$, it follows that
\begin{eqnarray}\label{random2}
&&V((m+1)h)\cr
&=&\delta^T(mh)\Phi^T((m+1)h,mh)\Phi((m+1)h,mh)\delta(mh)+(\tilde\xi_{m}^{mh})^T(\tilde\xi_{m}^{mh})\cr
&&+~2\delta^T(mh)\Phi^T((m+1)h,mh)\tilde\xi_{m}^{mh}\cr
&=&\delta^T(mh)\Big[\Phi^T((m+1)h,mh)\Phi((m+1)h,mh)-I_N\cr
&&+~\sum_{i=mh}^{(m+1)h-1}c(i)[P_NL_{\mathcal G(i)}+L_{\mathcal G(i)}^TP_N^T]\Big]\delta(mh)\cr
&&+~V(mh)-\delta^T(mh)\sum_{i=mh}^{(m+1)h-1}c(i)[P_NL_{\mathcal G(i)}+L_{\mathcal G(i)}^TP_N^T]\delta(mh)+(\tilde\xi_{m}^{mh})^T(\tilde\xi_{m}^{mh})\cr
&&+~2\delta^T(mh)\Phi^T((m+1)h,mh)\tilde\xi_{m}^{mh}.
\end{eqnarray}
Noting that $\delta(mh)\in\mathcal{F}_{\xi, \mathcal{A}}(mh-1)$, by the properties of conditional expectation, we know that
\begin{eqnarray}
\label{410add}
&&E\left[\delta^T(mh)\Phi^T((m+1)h,mh)\Phi((m+1)h,j+1)P_ND(j)Y(j)\xi(j)\right]\nonumber\\
&=&E\Big[\delta^T(mh)E\Big[\Phi^T((m+1)h,mh)\Phi((m+1)h,j+1)\cr
&&\times P_ND(j)Y(j)\xi(j)|{\mathcal F_{\xi,{\mathcal A}}}(j-1)\Big]\Big],\ mh\le j\le (m+1)h-1, m\ge0.
\end{eqnarray}
By Assumptions \textbf{(A2)}, \textbf{(A5)} and Lemma \ref{independence}, we have
\ban
&&E\left[\Phi^T((m+1)h,mh)\Phi((m+1)h,j+1)P_ND(j)Y(j)\xi(j)|{\mathcal F_{\xi,{\mathcal A}}}(j-1)\right]\cr
&=&E\left[\Phi^T((m+1)h,mh)\Phi((m+1)h,j+1)P_ND(j)|{\mathcal F_{\xi,{\mathcal A}}}(j-1)\right]\cr
&&\times Y(j)E[\xi(j)|\mathcal F_{\xi,{\mathcal A}}(j-1)]\cr
&=&E\left[\Phi^T((m+1)h,mh)\Phi((m+1)h,j+1)P_ND(j)|{\mathcal F_{{\mathcal A}}}(j-1)\right]\cr
&&\times Y(j)E[\xi(j)|{\mathcal F_{\xi}}(j-1)]\cr
&=&\textbf{0}_{N\times N}.
\ean
This together with (\ref{tildexi}) and (\ref{410add}) gives
\bna
\label{42}
E\left[\delta^T(mh)\Phi^T((m+1)h,mh)\tilde\xi_{m}^{mh}\right]=0.
\ena
By Assumptions \textbf{(A3)} and \textbf{(A4)}, we know that there exist positive integer $m_0$ and positive constant $C_{1}$, such that $c^2(mh)\le C_{1}c^2((m+1)h)$, $\forall$  $m\geq m_{0}$, and  $c(k)\le1$, $\forall$ $k\ge m_0h$.
By condition (b.2) and the conditional Lyapunov inequality, we obtain that
\bna
\label{conjensen}
\sup_{k\ge0}E[\|L_{\mathcal G(k)}\|^{i}|\mathcal F_{\mathcal A}(k-1)]\le\sup_{k\ge0}[E[\|L_{\mathcal G(k)}\|^{2^{h}}|\mathcal F_{\mathcal A}(k-1)]]^{\frac{i}{2^h}}\le \rho_{0}^i\ {\rm a.s.},\ \forall\ 2\le  i\le 2^{h}.
\ena
Denote the combinatorial number of choosing $i$ elements from $2h$ elements by $M_{2h}^i$. By termwise multiplication and using the H\"{o}lder  inequality repeatedly, noting that
$$E[\|L_{\mathcal{G}(k)}\|^{l}|\mathcal{F}_{\mathcal{A}}(mh-1)]= E[E[\|L_{\mathcal{G}(k)}\|^{l}|\mathcal{F}_{\mathcal{A}}(k-1)]|\mathcal{F}_{\mathcal{A}}(mh-1)],\ 2\le  l\le 2^{h},\ k\geq mh,$$ from (\ref{conjensen}), we have
\begin{eqnarray}\label{random31}
&&E\Big\{\Big\|\Phi^T((m+1)h,mh)\Phi((m+1)h,mh)
-I_N+\sum_{i=mh}^{(m+1)h-1}c(i)(P_NL_{\mathcal G(i)}+L_{\mathcal G(i)}^TP_N^T)
\Big\|\cr
&&~~~~\Big|\mathcal F_{\mathcal A}(mh-1)\Big\}\cr
&\le& \Big(C_{1}\sum_{i=2}^{2h}M_{2h}^i\rho_{0}^i\Big)c^2((m+1)h)\cr
&=&C_{1}[(1+\rho_{0})^{2h}-1-2h\rho_{0}]c^2((m+1)h),\ m\geq m_{0}.
%&&\le C_{1}\rho_{0}\sum_{l=2}^{2h}M_{2h}^l c^2((m+1)h),a.s.
\end{eqnarray}
Denote the symmetrized graph of  $\mathcal G(i|mh-1)$ by $\hat{\mathcal G}(i|mh-1)$, $mh \le i\le(m+1)h-1$. Noting that $ \mathcal G(i|i-1)$ is balanced, a.s.,  we know that $ \mathcal G(i|mh-1)$ is balanced, a.s. Then $E[\hat{L}_{\mathcal G(i)}|\mathcal F_{\mathcal A}(mh-1)]$ is the Laplacian matrix of $\hat{\mathcal G}(i|mh-1)$, a.s., $mh \le i\le(m+1)h-1$. So
$$
\sum_{i=mh}^{(m+1)h-1}E[ \hat{L}_{\mathcal G(i)}|\mathcal F_{\mathcal A}(mh-1)] \hbox{~is~the~Laplacian~matrix~of} \sum_{i=mh}^{(m+1)h-1}\hat{\mathcal G}(i|mh-1)\ {\rm a.s.} $$
Furthermore, by Assumption \textbf{(A5)} and Lemma \ref{independence}, we have
\ban
&&E\left[\delta^T(mh)\left[\sum_{i=mh}^{(m+1)h-1}c(i)(P_NL_{\mathcal G(i)}+L_{\mathcal G(i)}^TP_N^T)\right]\delta(mh)\right]\cr
&=&2E\left[\delta^T(mh)\left[\sum_{i=mh}^{(m+1)h-1}c(i)E[ \hat{L}_{\mathcal G(i)}|\mathcal F_{\xi,\mathcal{A}}(mh-1)]\right]\delta(mh)\right]\nonumber\\
&=&2E\left[\delta^T(mh)\left[\sum_{i=mh}^{(m+1)h-1}c(i)E[ \hat{L}_{\mathcal G(i)}|\mathcal F_{\mathcal A}(mh-1)]\right]\delta(mh)\right],
\ean
which together with Assumption \textbf{(A4)} and  condition (b.1) leads to
\bna\label{random4}
&&E\left[\delta^T(mh)\left[\sum_{i=mh}^{(m+1)h-1}c(i)(P_NL_{\mathcal G(i)}+L_{\mathcal G(i)}^TP_N^T)\right]\delta(mh)\right]\nonumber\\
%&=&2E\left\{\delta^T(mh)\left[\sum_{i=mh}^{(m+1)h-1}c(i)E[ \hat{L}_{\mathcal G(i)}|\mathcal F_{\mathcal A}(mh-1)]\right]\delta(mh)\right\}\nonumber\\
&\ge&2c((m+1)h)E\left[\delta^T(mh)\left[\sum_{i=mh}^{(m+1)h-1}E[ \hat{L}_{\mathcal G(i)}|\mathcal F_{\mathcal A}(mh-1)]\right]\delta(mh)\right]\cr
&\ge&2c((m+1)h)E\left[\lambda_m^{mh}V(mh)\right]\cr
&\ge&2c((m+1)h)E\left[\inf_{m\ge0}(\lambda_m^{mh})V(mh)\right] \cr
&\ge&2\theta c((m+1)h) E[V(mh)]\ \mbox{a.s.}
\ena
By  Assumptions \textbf{(A2)}, \textbf{(A5)} and Lemma \ref{independence},  it follows that
\ban
&&E[\xi^T(i)Y^T(i)D^T(i)P_N\Phi^T((m+1)h,i+1)\Phi((m+1)h,j+1)P_ND(j)Y(j)\xi(j)]\cr
&=&E[E[\xi^T(i)Y^T(i)D^T(i)P_N\Phi^T((m+1)h,i+1)\Phi((m+1)h,j+1)\mid\mathcal{F}_{\xi, \mathcal{A}}(j)]\cr
&&\times P_ND(j)Y(j)\xi(j)]\cr
&=&E[E[\xi^T(i)Y^T(i)\mid\mathcal{F}_{\xi, \mathcal{A}}(j)]\cr
&&\times E[D^T(i)P_N\Phi^T((m+1)h,i+1)\Phi((m+1)h,j+1)\mid\mathcal{F}_{\mathcal{A}}(j)]P_ND(j)Y(j)\xi(j)]\cr
&=&E[E[E[\xi^T(i)\mid\mathcal{F}_{\xi}(i-1)]Y^T(i)\mid\mathcal{F}_{\xi, \mathcal{A}}(j)]\cr
&&\times E[D^T(i)P_N\Phi^T((m+1)h,i+1)\Phi((m+1)h,j+1)\mid\mathcal{F}_{\mathcal{A}}(j)]P_ND(j)Y(j)\xi(j)]\cr
&=&0,\ i>j,
\ean
which together with the definition of $\tilde\xi_{m}^{mh}$ gives
\bna
\label{xitildeaddadd1}
&&E[(\tilde\xi_{m}^{mh})^T(\tilde\xi_{m}^{mh})]\cr
&=&\sum_{i=mh}^{(m+1)h-1}c^2(i)E[\xi^T(i)Y^T(i)D^T(i)P_N\Phi^T((m+1)h,i+1)\cr
&&\times\Phi((m+1)h,i+1)P_ND(i)Y(i)\xi(i)]\cr
&\leq&\sum_{i=mh}^{(m+1)h-1}c^2(i)E\{\|\Phi^T((m+1)h,i+1)\Phi((m+1)h,i+1)\|\|D^T(i)D(i)\|\cr
&&\times\|Y(i)\|^2\|\xi(i)\|^2\}\cr
&=&\sum_{i=mh}^{(m+1)h-1}c^2(i)E\{\|Y(i)\|^2E[\|\Phi^T((m+1)h,i+1)\Phi((m+1)h,i+1)\|\cr
&&\times\|D^T(i)D(i)\|\mid\mathcal{F}_{\mathcal{A}}(i-1)]E[\|\xi(i)\|^2|\mathcal{F}_{\xi}(i-1)]\}.
\ena
By condition (b.2), we know that there is a constant $\rho_1^{'}$, such that
\ban
\sup_{k\geq0}\left[E[\|D^T(k)D(k)\|^2|\mathcal F_{\mathcal A}(k-1)]\right]^{1/2}\leq\rho_1^{'}\ {\rm a.s.},
\ean
which together with the conditional H\"{o}lder inequality and Cr-inequality leads to
\ban
&&E\{\|\Phi^T((m+1)h,i+1)\Phi((m+1)h,i+1)\|\|D^T(i)D(i)\||\mathcal F_{\mathcal A}(i-1)\}\cr
%&=&E\{\|\Phi^T((m+1)h,i+1)\Phi((m+1)h,i+1)\|\max_{1\le i\le N}(\sum_{j=1}^N a_{ij}^2(k))|\mathcal F_{\mathcal A}(i-1)\}\cr
%&\le& E\{\|\Phi^T((m+1)h,i+1)\Phi((m+1)h,i+1)\|\|L_{\mathcal G(i)}\|^2|\mathcal F_{\mathcal A}(i-1)\}\cr
&\le& \rho_{1}^{'}\{E\{\|\Phi^T((m+1)h,i+1)\Phi((m+1)h,i+1)\|^2|\mathcal F_{\mathcal A}(i-1)\}\}^{\frac{1}{2}}\cr
&\le& \rho',\ mh \le i\le (m+1)h-1,\ m\ge m_0,
%&&\le \rho_{0}\sum_{l=1}^{2(h-1)}M_{2(h-1)}^l
\ean
where $\rho'=\rho_{1}^{'}\left\{\big(\sum_{j=0}^{2(h-1)}M_{2(h-1)}^j\big)\sum_{l=0}^{2(h-1)}M_{2(h-1)}^l\rho_{0}^{2l}\right\}^{\frac{1}{2}}$.
Then by  (\ref{noiseintensity}), (\ref{xitildeaddadd1}) and the above, we get
\bna\label{noiseestimate41}
&&E[(\tilde\xi_{m}^{mh})^T(\tilde\xi_{m}^{mh})]\cr
&\le &\rho'\sum_{i=mh}^{(m+1)h-1}c^2(i)E\{4\sigma^2V(i)E[\|\xi(i)\|^2\mid\mathcal{F}_{\xi}(i-1)]+2b^2E[\|\xi(i)\|^2\mid\mathcal{F}_{\xi}(i-1)]\}\cr
&\le&4\sigma^2\b\rho'\sum_{i=mh}^{(m+1)h-1}c^2(i)E[V(i)]+2b^2\b \rho'\sum_{i=mh}^{(m+1)h-1}c^2(i),\ m\ge m_0.
\ena
Finally, by (\ref{random2}), (\ref{42}), (\ref{random31}), (\ref{random4}) and (\ref{noiseestimate41}), we have
\bna\label{middleinequlity}
&&E[V((m+1)h)]\cr&\le&\left(1-2\theta c((m+1)h)+c^2((m+1)h)C_{1}[(1+\rho_{0})^{2h}-1-2h\rho_{0}] \right)E[V(mh)]\cr&&+4\sigma^2\b \rho'\sum_{i=mh}^{(m+1)h-1}c^2(i)E[V(i)]+2b^2\b \rho'\sum_{i=mh}^{(m+1)h-1}c^2(i),\ m\geq m_0.
\ena
We call (\ref{middleinequlity}) the difference inequality of stochastic Lyapunov function. This together with $\sup_{k\geq0}E[V(k)]<\infty$ (\textbf{Step} 1) and  (\ref{middleinequlity}) gives
\bna\label{middleinequlity11}
&&E[V((m+1)h)]\cr&\le&(1-2\theta c((m+1)h)+c^2((m+1)h) C_{1}[(1+\rho_{0})^{2h}-1-2h\rho_{0}] )E[V(mh)]\cr&&+C_{2}\sum_{i=mh}^{(m+1)h-1}c^2(i),\ m\ge m_0,
\ena
where $C_2=(4\sigma^2\sup_{k\ge0}E[V(k)]+2b^2)\b \rho'$.

By Assumption \textbf{(A3)},  we know that there exists positive integer $m_1$ such that
\bna\label{dkfjnjk}
0<2\theta c((m+1)h)- c^2((m+1)h)C_{1}[(1+\rho_{0})^{2h}-1-2h\rho_{0}]\le1,\forall~m\ge m_1
\ena
and
\bna\label{dsklfsalfj}
\sum_{m=0}^\infty\{2\theta c((m+1)h)- c^2((m+1)h) C_{1}[(1+\rho_{0})^{2h}-1-2h\rho_{0}]\}=\infty.
\ena
And by Assumption \textbf{(A4)}, we get
\bna\label{energy27}
\lim_{m\to\infty}\frac{C_{2}\sum_{i=mh}^{(m+1)h-1}c^2(i)}{2\theta c((m+1)h)- c^2((m+1)h) C_{1}[(1+\rho_{0})^{2h}-1-2h\rho_{0}]}=0.%\sum_{j=mh}^{(m+1)h-1}c^2(j)=o(c((m+1)h)),m\to\infty.
\ena
Then by Lemma~\ref{lemma1} and (\ref{middleinequlity11})-(\ref{energy27}), we get $E[V(mh)]\to0,m\to\infty$.
Thus, for any given $\epsilon>0$, there exists positive integer $m_2$ such that $E[V(mh)]<\epsilon$, $m\ge m_2,$ and $\sum_{i=m_2h}^\infty c^2(i)<\epsilon$. Let $m_k=\lfloor\frac{k}{m_2}\rfloor$. Then for any given $k\ge m_2h$, we have  $m_k\ge m_2$ and  $0\le k-m_kh\le h$. Therefore, by (\ref{energy28}), we have
\ban
E[V(k+1)]
&\le&\prod_{i=m_kh}^k[1+c^2(i)(\rho_{0}^2+4\rho_{1}\beta\sigma^2)]E[V(m_kh)]\cr&&+2\rho_{1} b^2\b\sum_{i=m_kh}^k \prod_{j=i+1}^k[1+c^2(j)(\rho_{0}^2+4\rho_{1}\beta\sigma^2)]c^2(i)\cr
%由$c(k)\downarrow0$, 可知存在常数$C_{6}$ 使得
&\le& \exp((\rho_{0}^2+4\rho_{1}\beta\sigma^2)\sum_{i=0}^\infty c^2(i))(1+2\rho_{1} b^2\b)\epsilon,\ k\ge m_2h.
\ean
where $\prod_{j=k+1}^{k}[1+(\rho_{0}^2+4\rho_{1}\beta\sigma^2)c^2(j)]$ is defined as $1$. Then by the arbitrariness of $\epsilon$, we get
\bna\label{evk} E[V(k)]\to0,k\to\infty.\ena

\noindent
\textbf{Step~3}: To prove $\{\frac{1}{N}\sum_{i=1}^Nx_i(k)$, $k\ge0\}$ converges in mean square and almost surely.

Let $\tilde L_{\mathcal G(k)}=L_{\mathcal G(k)}-E[ L_{\mathcal G(k)}|\mathcal F_{\mathcal A}(k-1)]$, $k\ge0$.
Noting that the associated digraph of Laplacian matrix $E[ L_{\mathcal G(k)}|\mathcal F_{\mathcal A}(k-1)]$ is balanced a.s., we know that $\textbf{1}^TE[ L_{\mathcal G(k)}|\mathcal F_{\mathcal A}(k-1)]=\textbf{0}_N^T$\ a.s. Left multiplied by $\frac{1}{N}\textbf{1}_N^T$ on both sides of (\ref{driven2}), and then making a summation from $0$ to $n-1$ with respect to $k$, we have
\bna\label{sfkjsdf} \frac{1}{N}\sum_{j=1}^{N}x_j(n)&=&\frac{1}{N}\sum_{j=1}^{N}x_j(0)-\frac{1}{N}\textbf{1}^T\sum_{k=0}^{n-1}c(k)
L_{\mathcal G(k)}X(k)+\frac{1}{N}\textbf{1}^T\sum_{k=0}^{n-1}c(k)D(k)Y(k)\xi(k)\cr
&=&\frac{1}{N}\sum_{j=1}^{N}x_j(0)-\frac{1}{N}\textbf{1}^T\sum_{k=0}^{n-1}c(k)
\tilde L_{\mathcal G(k)}X(k)\cr
&&+\frac{1}{N}\textbf{1}^T\sum_{k=0}^{n-1}c(k)D(k)Y(k)\xi(k).
\ena
Noting that
\begin{eqnarray*}\label{martingaleproof10}
&&E[\tilde L_{\mathcal G(m+i)}X(m+i)|{\mathcal F_{\xi,{\mathcal A}}}(m)]\cr
&=&E\{E[\tilde L_{\mathcal G(m+i)}X(m+i)|{\mathcal F_{\xi,{\mathcal A}}}(m)]|{\mathcal F_{\xi,{\mathcal A}}}(m+i-1)\} \cr
&=&E\{E[\tilde L_{\mathcal G(m+i)}X(m+i)|{\mathcal F_{\xi,{\mathcal A}}}(m+i-1)]|{\mathcal F_{\xi,{\mathcal A}}}(m)\}\cr
&=&E\{E [\tilde L_{\mathcal G(m+i)} |{\mathcal F_{\xi,{\mathcal A}}}(m+i-1)]X(m+i)|{\mathcal F_{\xi,{\mathcal A}}}(m)\},1\le i\le n-m-1,
\end{eqnarray*}
by  the definition of $\tilde L_{\mathcal G(k)}$ and Assumption \textbf{(A5)},  it is known that $E [\tilde L_{\mathcal G(k)} |{\mathcal F_{\xi,{\mathcal A}}}(k-1)]=E [\tilde L_{\mathcal G(k)} |{\mathcal F_{{\mathcal A}}}(k-1)]=O_{N\times N}$, $k\ge0$. Thus, from the above equality, we get
$$
E[\tilde L_{\mathcal G(m+i)}X(m+i)|{\mathcal F_{\xi,{\mathcal A}}}(m)]=\textbf{0}_N,1\le i\le n-m-1,
$$
which gives
\begin{eqnarray*}\label{martingaleproof}
&&E\Bigg[\sum_{k=0}^{n-1}\tilde L_{\mathcal G(k)}X(k)\bigg|{\mathcal F_{\xi,{\mathcal A}}}(m)\Bigg]\cr
&=&E\Bigg[\sum_{i=0}^m\tilde L_{\mathcal G(i)}X(i)\bigg|
{\mathcal F_{\xi,{\mathcal A}}}(m)\Bigg]+E\Bigg[\sum_{i=m+1}^{n-1}\tilde L_{\mathcal G(i)}X(i)\bigg|{\mathcal F_{\xi,{\mathcal A}}}(m)\Bigg]\cr
&=&E\Bigg[\sum_{i=0}^m\tilde L_{\mathcal G(i)}X(i)\bigg|
{\mathcal F_{\xi,{\mathcal A}}}(m)\Bigg],\ \forall\ m< n-1.
%&&=\tilde L_{\mathcal G(0)}X(0)+...+\tilde L_{\mathcal G(m)}X(m)
\end{eqnarray*}
%\sum_{k=0}^{m}\tilde L_{\mathcal G(k)}X(k),
Then by the above and the  definition of martingales, we know that $\Big\{\frac{1}{N}\textbf{1}_N^T
\sum_{k=0}^{n}c(k)\tilde L_{\mathcal G(k)}X(k)$, ${\mathcal F_{\xi,{\mathcal A}}}(n)$, $n\ge0\Big\}$ is a martingale.
On the other hand, by  (\ref{sfkjsdf}), we know that
\bna
\label{asmfhlaskd}
&&\sup_{n\ge0}E\left\|\sum_{k=0}^{n-1}c(k)\tilde L_{\mathcal G(k)}X(k)\right\|^2\cr
&\le&\sup_{n\ge0}\sum_{k=0}^{n-1}c^2(k)E[\|X(k)\|^2\|\tilde L_{\mathcal G(k)}\|^2]\cr
&\le&\sup_{k\ge0}E[\|\tilde L_{\mathcal G(k)}\|^2|\mathcal F_{{\mathcal A}}(k-1)]\sup_{k\ge0}E\|X(k)\|^2\sum_{k=0}^{\infty}c^2(k).
\ena
By condition (b.2), we know that
\bna\label{safkjh}
\sup_{k\ge0}E[\|\tilde L_{\mathcal G(k)}\|^2|\mathcal F_{{\mathcal A}}(k-1)]<\infty\ {\rm a.s.}
\ena
From (\ref{driven2}), (\ref{dafmalka}) and condition (b.2), we get
\bna\label{sfdgs}
&&E[\|X(k+1)\|^2]\cr&=&E[X^T(k)(I_N-c(k)L_{\mathcal G(k)}^T)(I_N-c(k)L_{\mathcal G(k)})X(k)]\cr
&&+c^2(k)E[\xi^T(k)Y^T(k)D^T(k)D(k)Y(k)\xi(k)]\cr
&\le& E[\|X(k)\|^2]+c^2(k)E[\|X(k)\|^2\|L_{\mathcal G(k)}\|^2]+c^2(k)E[\|Y(k)\|^2\|\xi(k)\|^2\|D^T(k)D(k)\|]\cr
&\le& E[\|X(k)\|^2]+c^2(k)\rho_{0}^2 E[\|X(k)\|^2]+c^2(k)\b \rho_{1} E[4\sigma^2V(k)+2b^2]\cr
&\le&(1+c^2(k)\rho_{0}^2)E[\|X(k)\|^2]+\b \rho_{1}(4\sigma^2\sup_{k\ge0}E[V(k)]+2b^2)c^2(k),
\ena
which together with Lemma \ref{robbins} and Assumption \textbf{(A3)} gives
%\bna\label{askjsfh}
$\sup_{k \ge0}E[\|X(k)\|^2]<\infty$.
%\ena
Then by (\ref{asmfhlaskd}) and (\ref{safkjh}), we know that
\ban\label{sss}\sup_{n\ge0}E\left\|\sum_{k=0}^{n-1}c(k)\tilde L_{\mathcal G(k)}X(k)\right\|^2<\infty.\ean
This together with Lemma \ref{aksfmo} leads to
\bna\label{sum}
\frac{1}{N}\textbf{1}_N^T\sum_{k=0}^{n-1}c(k)\tilde L_{\mathcal G(k)}X(k)\ {\rm converges\ a.s.\ and\ in}\ \mathfrak{L}_2.\ena
From Assumptions \textbf{(A2)} and \textbf{(A5)}, it follows that
\begin{eqnarray*}\label{martingaleproof1}
&&E\left[\sum_{k=0}^{n-1}c(k)D(k)Y(k)\xi(k)|\mathcal F_{\xi,{\mathcal A}}(j)\right]\cr
&=&\sum_{k=0}^{j}c(k)D(k)Y(k)\xi(k)+\sum_{k=j+1}^{n-1}E\left[E(c(k)D(k)Y(k)\xi(k)|\mathcal F_{\xi,{\mathcal A}}(k-1))|\mathcal F_{\xi,{\mathcal A}}(j)\right]\nonumber\\
&=&\sum_{k=0}^{j}c(k)D(k)Y(k)\xi(k),\forall ~j< n-1.
\end{eqnarray*}
Thus, the adaptive sequence $\big\{\sum_{j=0}^{n}c(k)D_{\mathcal G(k)}Y(k)\xi(k)$, $\mathcal F_{\xi,{\mathcal A}}(n),n\ge0\big\}$ is a martingale.
Then by (\ref{noiseintensity}) and  condition (b.2), we have
\begin{eqnarray*}\label{convergence1}
&&\sup_{n\ge0}E\left\|\sum_{k=0}^{n-1}c(k)D(k)Y(k)\xi(k)\right\|^2\cr
&=&\sup_{k\ge0}E\left[\Big(\sum_{k=0}^{n-1}c(k)\xi^T(k)Y^T(k)D^T(k)\Big)\Big(\sum_{k=0}^{n-1}c(k)D(k)Y(k)\xi(k)\Big)\right]\cr
&=&\sup_{k\ge0}\sum_{k=0}^{n-1}E\left[c^2(k)\xi^T(k)Y^T(k)D^T(k)D(k)Y(k)\xi(k)\right]\cr
&\le&\beta\sup_{k\ge0}E[\|D^T(k)D(k)
\||\mathcal F_{\mathcal A}(k-1)]\sup_{n\ge0}\sum_{k=0}^{n-1}c^2(k)
E\|Y(k)\|^2\nonumber\\
&\le&\beta\rho_{1}\sup_{n\ge0}\sum_{k=0}^{n-1} c^2(k)(4\sigma^2 E[V(k)]+2b^2).
\end{eqnarray*}
By Assumption \textbf{(A3)}, the boundedness of $E[V(k)]$ and the above, we get
\bna\label{convergence2}
\sup_{n\ge0}E\left\|\sum_{k=0}^{n-1}c(k)D(k)Y(k)\xi(k)\right\|^2<\infty,
\ena
which together with Lemma \ref{aksfmo} gives
\bna\label{sajdhkasdnaj}
\frac{1}{N}\textbf{1}_N^T\sum_{k=0}^{n-1}c(k)D(k)Y(k)\xi(k)\ {\rm converges},\ k\to\infty\ {\rm a.s.\ and\ in}\ \mathfrak{L}_2.
\ena
Finally, by (\ref{sfkjsdf}), (\ref{sum}) and (\ref{sajdhkasdnaj}) we know that
\bna\label{ksmf}\frac{1}{N}\sum_{j=1}^{N}x_j(n)\longrightarrow x^*,\ n\to\infty\ {\rm a.s.\ and\ in}\ \mathfrak{L}_2,\ena
where
\bna\label{random3}x^*=\frac{1}{N}\sum_{j=1}^{N}x_j(0)-\frac{1}{N}
\textbf{1}_N^T\sum_{k=0}^{\infty}
c(k)\tilde L_{\mathcal G(k)}X(k)+\frac{1}{N}\textbf{1}_N^T\sum_{k=0}^{\infty}c(k)D(k)Y(k)\xi(k).\ena

\noindent
\textbf{Step~4}: To prove all $x_i(k)$, $i\in\mathcal V$ converge to $x^*$ as $k\to\infty$ in mean square and almost surely.

By the  definition of $V(k)$, %it is known that
%\bna
%\label{snjf}V(k)=\sum_{i=1}^N\Big(x_i(k)-\frac{\sum_{i=1}^Nx_i(k)}{N}\Big)
%\ena
(\ref{evk}) and (\ref{ksmf}), we have
$$x_i(k)\longrightarrow x ^*,\ k\to\infty,\ {\rm in}\ \mathfrak{L}_2,\ i\in\mathcal V.$$
Taking conditional expectation on both sides of (\ref{dgks}) gives
\ban
E[V(k+1)|\mathcal F_{\xi,\mathcal A}(k-1)]
\le V(k)[1+c^2(k)(\rho_{0}^2+4\sigma^2 \rho_{1}\b)]+2b^2\rho_{1}\b c^2(k).
\ean
Then by Lemma \ref{robbins} and Assumption \textbf{(A3)},  we obtain
\ban
V(k)\to {\rm a\ finite\ random\ variable},\ k\to\infty\  {\rm a.s.},
\ean
which together with (\ref{evk}) gives
\ban
V(k)\to 0,\ k\to\infty\ {\rm a.s.}.
\ean
Then by (\ref{ksmf}), we have
$$x_i(k)\longrightarrow x ^*,\ k\to\infty\ {\rm a.s.},\ i\in\mathcal V.$$

\noindent
\textbf{Step 5}: To compute the mathematical expectation of $x^*$.

By (\ref{sum}), the definition of $\tilde L_{\mathcal G(k)}$ and Assumption \textbf{(A5)}, we have
$$E\left[\frac{1}{N}
\textbf{1}_N^T\sum_{k=0}^{\infty}
c(k)\tilde L_{\mathcal G(k)}X(k)\right]=\lim_{n\to\infty}E\left[\frac{1}{N}
\textbf{1}_N^T\sum_{k=0}^{n-1}
c(k)\tilde L_{\mathcal G(k)}X(k)\right]=0.$$
Similarly, by Assumptions \textbf{(A2)} and \textbf{(A5)}, we have
$$
E\left[\frac{1}{N}\textbf{1}_N^T\sum_{k=0}^{\infty}c(k)D(k)Y(k)\xi(k)\right]=\lim_{n\to\infty}E\left[\frac{1}{N}
\textbf{1}_N^T\sum_{k=0}^{n-1}c(k)D(k)Y(k)\xi(k)\right]=0.
$$
This together with (\ref{random3}) gives
\bna
\label{expect1addadd1}
E(x^*)=\frac{1}{N}\sum_{j=1}^{N}x_j(0).
\ena

\noindent
\textbf{Step 6}: To estimate the variance of $x^*$.

From (\ref{energy28}), by iterative calculations, we have
\bna\label{akfnsasdkn}
E[V(k+1)]&\le&\prod_{i=0}^{k}[1+(\rho_{0}^2+4\beta\sigma^2\rho_{1})c^2(i)]V(0)\cr
&&~+2\rho_{1} b^2\b \sum_{i=0}^{k}c^2(i)\prod_{j=i+1}^{k}[1+(\rho_{0}^2+4\beta\sigma^2\rho_{1})c^2(j)],
\ena
where $\prod_{j=k+1}^{k}[1+(\rho_{0}^2+4\beta\sigma^2\rho_{1})c^2(j)]=1$.
Actually, for $\forall ~k\ge j$,
$$
\prod_{i=j}^{k}(1+(\rho_{0}^2+4\beta\sigma^2\rho_{1})c^2(i))\le\exp\Big((\rho_{0}^2+4\beta\sigma^2\rho_{1})\sum_{i=j}^kc^2(i)\Big)\le
\exp\Big((\rho_{0}^2+4\beta\sigma^2\rho_{1})\sum_{i=0}^\infty c^2(i)\Big).
$$
This together with (\ref{akfnsasdkn}) leads to
\bna\label{lkjfla}
\sup_{k\geq0}E[V(k)]\le\exp\Big((\rho_{0}^2+4\beta\sigma^2\rho_{1})\sum_{i=0}^\infty c^2(i)\Big)\Big\{V(0)+2\rho_{1} b^2\b \sum_{i=0}^{\infty}c^2(i)\Big\}= q_v.
\ena
Similarly, by (\ref{sfdgs}) and the above, we have
\bna\label{statebound}
 E\|X(k+1)\|^2&\le&(1+c^2(k)\rho_{0}^2)E\|X(k)\|^2+\b \rho_{1}(4\sigma^2 q_v+2b^2)c^2(k)\cr
 &\le& \exp\Big(\rho_{0}^2\sum_{k=0}^\infty c^2(k)\Big)\Big\{\|X(0)\|^2+\b \rho_{1}(4\sigma^2 q_v+2b^2)\sum_{k=0}^{\infty}c^2(k)\Big\}\cr
 &=& q_x.
\ena
Then by  (\ref{sum}), (\ref{sajdhkasdnaj}), (\ref{random3}), (\ref{expect1addadd1}), the dominated convergence theorem and Cr-inequality, we have
\bna\label{saljoa}
Var(x^*)&=&E\Big[\frac{1}{N}\textbf{1}_N^T\sum_{k=0}^{\infty}c(k)D(k)Y(k)\xi(k)-\frac{1}{N}\textbf{1}_N^T\sum_{k=0}^{\infty}c(k)\tilde L_{\mathcal
G(k)}X(k)\Big]^2\cr
&\le& 2E\Big[\frac{1}{N}\textbf{1}_N^T\sum_{k=0}^{\infty}c(k)D(k)Y(k)\xi(k)\Big]^2+2E\Big[\frac{1}{N}\textbf{1}_N^T\sum_{k=0}^{\infty}c(k)\tilde L_{\mathcal
G(k)}X(k)\Big]^2\cr
&\le&2\lim_{n\to\infty}E\Big[\frac{1}{N}\textbf{1}_N^T\sum_{k=0}^{n-1}c(k)\tilde L_{\mathcal
G(k)}X(k)\Big]^2\cr
&&+2\lim_{n\to\infty}E\Big[\frac{1}{N}\textbf{1}_N^T\sum_{k=0}^{n-1}c(k)D(k)Y(k)\xi(k)\Big]^2.
\ena
For the first term on right hand side of (\ref{saljoa}), by the definition of $\tilde L_{\mathcal
G(k)}$, Assumption \textbf{(A5)}, Cr-inequality and (\ref{statebound}), we have
\bna\label{asffahkasd}
&&\lim_{n\to\infty}E\left[\frac{1}{N}\textbf{1}_N^T\sum_{k=0}^{n-1}c(k)\tilde L_{\mathcal
G(k)}X(k)\right]^2\cr
&=&\lim_{n\to\infty}E\left[\frac{1}{N}\textbf{1}_N^T\sum_{k=0}^{n-1}c(k) L_{\mathcal
G(k)}X(k)\right]^2\cr
&=&\frac{1}{N^2}\lim_{n\to\infty}\sum_{k=0}^{n-1}\left\{c^2(k)E\left[\sum_{i=1}^N x_i(k)\Big(\sum_{j=1}^Na_{ij}(k)-\sum_{j=1}^Na_{ji}(k)\Big)\right]^2\right\}\cr
&\le& \frac{1}{N}\lim_{n\to\infty}\sum_{k=0}^{n-1}\left\{c^2(k)\sum_{i=1}^N E\left[x_i^2(k)\Big(\sum_{j=1}^Na_{ij}(k)-\sum_{j=1}^Na_{ji}(k)\Big)^2\right]\right\}\cr
&\le& \frac{\rho_{2}}{N}\sum_{k=0}^{\infty}c^2(k)E\|X(k)\|^2\le\frac{\rho_{2}q_x}{N}\sum_{k=0}^{\infty}c^2(k).
\ena
For the second term, by Assumption \textbf{(A2)},  direct calculations gives
\bna\label{0011}
&&\lim_{n\to\infty}E\left[\frac{1}{N}\textbf{1}_N^T\sum_{k=0}^{n-1}c(k)D(k)Y(k)\xi(k)\right]^2\cr
&=&\frac{1}{N^2}\lim_{n\to\infty}E\left[\sum_{k=0}^{n-1}(\textbf{1}_N^Tc(k)D(k)Y(k)\xi(k))^2\right]\cr
&\le&\frac{1}{N^2}\lim_{n\to\infty}\sum_{k=0}^{n-1}c^2(k)E\left[\sum_{1\le i,j\le N}\xi_{ji}(k)a_{ij}(k)(\sigma_{ji}(x_j(k)-x_i(k))+b_{ji})\right]^2.
\ena
Then by Assumptions \textbf{(A5)}, condition (b.2) and Cr-inequality,
%and the equality
%$$\sum_{1\le i,j\le N}(x_j(k)-x_i(k))^2=2N\sum_{i=1}^N\left(x_i-\frac{\sum_{j=1}^Nx_j(k)}{N}\right)^2=2NV(k),$$
we have
\bna\label{011}
&&\lim_{n\to\infty}E\Big[\frac{1}{N}\textbf{1}_N^T\sum_{k=0}^{n-1}c(k)D(k)Y(k)\xi(k)\Big]^2\cr
%&=&\frac{1}{N^2}\lim_{n\to\infty}E\left[\sum_{k=0}^{n-1}(\textbf{1}_N^Tc(k)D(k)Y(k)\xi(k))^2\right]\cr
%&=&\frac{1}{N^2}\lim_{n\to\infty}E\Big[\sum_{k=0}^{n-1}\sum_{i=1}^{N}\sum_{j=1}^{N}c(k)a_{ij}(k)f_{ji}(k)\xi_{ji}(k)\Big]^2\cr
%&=&\frac{1}{N^2}\sum_{k=0}^{\infty}E\left\{\Big[\sum_{i=1}^{N}\sum_{j=1}^{N}c(k)a_{ij}(k)f_{ji}(k)\xi_{ji}(k)\Big]^2\right\}\cr
&\le& \frac{1}{N^2}\sum_{k=0}^\infty \left\{c^2(k)\sum_{(i,j)\in\mathcal E_{\mathcal G(k)}}E\Big[|\mathcal E_{\mathcal G(k)}|\xi_{ji}^2(k)a_{ij}^2(k)(\sigma_{ji}(x_j(k)-x_i(k))+b_{ji})^2\Big]\right\}\cr
&\le& \frac{2}{N^2}\sum_{k=0}^\infty \left\{c^2(k)\sum_{(i,j)\in\mathcal E_{\mathcal G(k)}}E\Big[|\mathcal E_{\mathcal G(k)}|\xi_{ji}^2(k)a_{ij}^2(k)(\sigma_{ji}^2(x_j(k)-x_i(k))^2+b_{ji}^2)\Big]\right\}\cr
&\le&\frac{2\b b^2\rho_{1}}{N^2}\sum_{k=0}^\infty c^2(k)+\frac{4\b\sigma^2 \rho_{1}}{N^2}\sum_{k=0}^{\infty} E[V(k)]c^2(k),
%&\le&\frac{2}{N^2}\b \rho_{1}\Big(b^2+2\sigma^2 q_v\Big)\sum_{i=0}^\infty c^2(i).
\ena
which together with (\ref{saljoa}) and (\ref{asffahkasd}) gives (\ref{013}). \qed

\vskip 0.2cm

\begin{remark}{\rm Most of existing literature on consensus-based distributed algorithms assumed that the edge weights, i.e., the entries of $\mathcal A_{\mathcal G(k)}$,  are nonnegative. In this paper, we only assume that the entries of $E[\mathcal A_{\mathcal G(k)}|\mathcal F_{\mathcal A}(k-1)]$ are nonnegative almost surely. This relaxation makes the algorithm more flexible at
the price of more difficult analysis, since $L_{\mathcal G(k)}$ is not a Laplacian matrix any more and some properties of Laplacian matrices
are not applicable.}
\end{remark}

\vskip 0.2cm

\begin{remark}{\rm Here,  Assumption \textbf{(A5)} requires that the graph flow and the measurement noises are mutually independent. And different from the most existing works on distributed averaging under random network graphs, here, neither the graph flow nor the process of measurement noises is required to be spatially or temporally independent. For the case with time-invariant random graphs, Porfiri and Stilwell \cite{047}  and Hatano and Mesbahi \cite{060} assumed independent channels.
For the case with  time-varying random graphs,  Boyd $et~al.$ \cite{051}, Kar and Moura \cite{059}, Tahbaz-Salehi and Jadbabaie \cite{061} and Long $et~al.$ \cite{022} assumed that $\{{\mathcal G(k)}$, $k\ge0\}$
is a sequence of independent random graphs. These spatial or temporal
independency requirements can not be always satisfied for real networks. Take a sensor network as the example. On the spatial scale, if a sensor node fails due to battery exhausted, then all channels between this node and its neighbours become inactive. This would happen randomly and the statistics of channels associated with this node are obviously spatially dependent.
On the temporal scale, the unreliability of channels would increase due to aging of sensors as time goes on. Thus, the statistics of channels are also temporally dependent. In this paper,  we do not require the spatial and temporal independency of the network graphs, which can cover more practical cases. Furthermore, we assume that the overall noises constitute a martingale difference sequence without requiring that the noises are spatial-temporal-independent as in \cite{021}-\cite{020} and \cite{0240}-\cite{023}). }
\end{remark}

\vskip 0.2cm

\begin{remark}
{\rm In \cite{0250}, the closed-loop
system is described by $x(t+1)=A(t)x(t)+B(t)m(t)$,
where $\{x(s):s\leq t\}$ is independent of $A(t)$, $B(t)$ and $m(t)$ for all $t\geq0$; and the disturbance process $m(t)$ is independent of $B(t)$.
This assumption  obviously fails for our model (\ref{driven2}).}
\end{remark}

\vskip 0.2cm

\begin{remark} {\rm We call condition (b.1) $\inf_{m\ge0}{{\lambda}}_{mh}^h\ge\theta$ a.s. the \emph{uniformly conditionally jointly connected condition}, i.e., the conditional digraphs $\mathcal G(k|k-1)$ are jointly connected over the intervals $[mh,(m+1)h-1]$, $m\ge0$, and the average algebraic connectivity is uniformly positive bounded away from zero.}
\end{remark}

\vskip 0.2cm
\begin{remark} {\rm  The inequality (\ref{013}) gives a upper bound of the  mean square steady-state error. There are three terms on the right hand side of (\ref{013}), which reflect the impacts of additive noises, multiplicative noises and the instantaneous unbalance of network graph on the final steady-state error, respectively. If the network graph is instantaneously balanced, i.e.,  $\sum_{j=1}^Na_{ij}(k)=\sum_{j=1}^Na_{ji}(k)$, $i=1,2,..., N$, a.s., then the third term vanishes. Especially, if the measurement noise sequence $\{\xi_{ji}(k), k=0,1,..., i,j=1,2...,N\}$ are both spatially and temporally independent, then from (\ref{0011}), we get
\bna
\label{varianceestiamteaddadd1}
Var(x^*)\le\frac{4c\b b^2\overline{\rho}_{1}}{N^2}+\frac{8\widetilde{c}\b\sigma^2\overline{\rho}_{1} }{N^2},
\ena
where $\overline{\rho}_{1}$ is a positive constant satisfying $\sup_{k\ge0}\max_{1\le i, j\le N}E\left[a_{ij}^2(k)|\mathcal F_{\mathcal A}(k-1)\right]\le\overline{\rho}_{1}$, a.s. Moreover, if $\beta=O(N)$ and $\overline{\rho}_{1}=O(1)$ as $N\to\infty$,
then $Var(x^*) = O(1/N)$,  $N\to\infty$, which means that the
larger the number of sensors is, the higher the accuracy of
information fusion is. At the same time, a sensor network
with large number of nodes is definitely uneconomic, so
there is a trade-off between the performance of the estimation
and the cost of the system for selecting the number of nodes.}
\end{remark}

\vskip 0.2cm
\begin{remark} {\rm The constant $\widetilde{c}$ in  (\ref{013}) and  (\ref{varianceestiamteaddadd1}) can be replaced by $q_vc$ from the estimation (\ref{lkjfla}). This removes the term $E[V(k)]$ in $\widetilde{c}$, however, makes the upper bound of the  mean square steady-state error more conservative.}
\end{remark}

%This tells us that the  mean square steady-state error is inversely proportional to the number of agents, which means that in some practical applications, such as the information fusion over wireless sensor networks, the larger the number of nodes is, the higher the estimate accuracy is. However, in the meantime,  a sensor network with large number of nodes is definitely
%uneconomic, so there would be a trade-off between the performance of the estimation and the cost of the system for selecting the
%number of nodes.

\section{Special cases}

%In Theorem \ref{mainresults}, the random graph flow is a general random process.
In this section, we consider two special classes of random graph flows:
(i) $\{\mathcal G(k),k\ge0\}$ is a Markov chain with countable state space; (ii) $\{\mathcal G(k),k\ge0\}$ is an independent process with uncountable state space. By the method of stochastic Lyapunov function based on random graph flows, we obtain sufficient conditions for mean square and almost sure average consensus. For these two special cases, condition (b.1) of Theorem \ref{mainresults} becomes more intuitive and condition (b.2) is weakened.

\subsection{Markovian switching flow}

\begin{definition}(\cite{053})
A Markov chain on a countable state space $\mathcal S$ with a stationary distribution $\pi$, and transition probability function $\mathbb P(x,\cdot)$  is called uniformly ergodic, if there exist positive constants $r>1$ and $R$ such that for all $x\in\mathcal S$,
$$
\|\mathbb P^n(x,\cdot)-\pi\|_1\le Rr^{-n}.
$$
Here, $\|\mathbb P^n(x,\cdot)-\pi\|_1=\sum_{y\in \mathcal{S}}|\mathbb P^n(x,y)-\pi(y)|$.% where $\mu,\nu$ are measures on $\mathcal B(\mathcal S)$
\end{definition}

\vskip 0.2cm
Denote $S_1=\{A_j, j=1, 2,...\}$, which is a countable set of generalized weighted adjacency matrices and the associated generalized Laplacian matrix of $A_j$ by $L_j$. Let $\hat L_{j}=\frac{L_j+L_j^T}{2}$.
%$S_2=\{ L_1, L_2, L_3,...\}$ and $S_3=\{\hat L_1,\hat L_2,\hat L_3,...\}$, which are  the countable state spaces of $\{\mathcal A_{\mathcal %G(k)},k\ge0\}$, $\{L_{\mathcal G(k)},k\ge0\}$ and $\{\hat L_{\mathcal G(k)},k\ge0\}$, respectively.
In this subsection, we consider the class of random graph flows as below, each element of which is a homogeneous and uniformly ergodic Markov chain with countable states and unique stationary distribution, i.e.
\ban
&&\Gamma_2=\Big\{\{\mathcal G(k),k\ge0\}|\{\mathcal A_{\mathcal G(k)},\ k\ge0\}\subseteq S_{1}, \mbox{and is a homogeneous and uniformly ergodic }\cr
&&~~~~~~~~\mbox{Markov chain with unique stationary
distribution}~\pi;  ~E[\mathcal A_{\mathcal G(k)}|\mathcal A_{\mathcal G(k-1)}]\succeq O_{N\times N}, ~\mbox{a.s.,} \cr
&&~~~~~~~~ \mbox{and the associated digraph of}\ E[\mathcal A_{\mathcal G(k)}|\mathcal A_{\mathcal G(k-1)}] ~\mbox{is balanced a.s.},\  k\ge0\Big\}.
\ean
Here, $\pi=[\pi_1,\pi_2,...]^T$, $\pi_{j}\geq0$, $\sum_{j=1}^{\infty}\pi_{j}=1$, where $\pi_{j}$ denotes $\pi(A_j)$.

We have the following theorem.

\begin{theorem}
\label{theoremmarkovadd1}
For systems (\ref{discretemodel})-(\ref{meaequa}) and the associated random graph flow $\{\mathcal G(k),k\ge0\}\in\Gamma_2$, assume that

(i) Assumptions \textbf{(A1)}-\textbf{(A5)} hold;

(ii) the associated graph of the Laplacian matrix $\sum_{j=1}^\infty\pi_j L_j$ contains a spanning tree;

(iii) $\sup_{j\ge1}\|\hat L_{j}\|<\infty$.

Then systems (\ref{discretemodel})-(\ref{meaequa}) achieve mean square and almost sure average consensus.
\end{theorem}

\noindent
\textbf{Proof}: Since $\{\mathcal A_{\mathcal G(k)}$, $k\ge0\}$ is a Markov chain, by the Markov property, we know that
$E[\mathcal A_{\mathcal G(k)}|\mathcal F_{\mathcal A}(k-1)]=E[\mathcal A_{\mathcal G(k)}|\mathcal A_{\mathcal G(k-1)}]$. Thus, $\{\mathcal G(k),k\ge0\}\in\Gamma_1$.

By the one-to-one correspondence among $\mathcal A_{\mathcal G(k)}$, $L_{\mathcal G(k)}$ and $\hat L_{\mathcal G(k)}$, we know that $\{ L_{\mathcal G(k)},k\ge0\}$ and $\{\hat L_{\mathcal G(k)},k\ge0\}$ are both homogeneous and uniformly ergodic Markov chains with the unique stationary
distribution $\pi$, whose state spaces are $S_2=\{ L_1, L_2, L_3,...\}$ and $S_3=\{\hat L_1,\hat L_2,\hat L_3,...\}$, respectively.
%where $L_j$ is the generalized Laplacian matrix corresponding to $A_j$, and $\hat L_j=\frac{L_j+L_j^T}{2}$, $j=1,2,...$
From (\ref{uniformalgebriacconnectivityaddadd1}), we know that
\bna\label{askfjak}
\lambda_{mh}^h&=&\lambda_2\left\{\sum_{i=mh}^{mh+h-1}E[\hat {L}_{\mathcal G(i)}|\hat L_{\mathcal G(mh-1)}=\hat L_{0}]\right\}\cr
&=&\lambda_2\left\{\sum_{i=1}^h\sum_{j=1}^\infty \hat {L}_j\mathbb P^i(\hat L_{0},\hat L_j)\right\},\ \forall\ \hat L_{0}\in S_3,\ \forall\ m\ge0, \ h\geq1.
\ena
Noting the uniform ergodicity of $\{\hat L_{\mathcal G(k)},k\ge0\}$ and the uniqueness of the stationary
distribution $\pi$, by condition (iii), we have
\ban
&&\left\|\frac{\sum_{i=1}^h\sum_{j=1}^\infty\hat {L}_j\mathbb P^i(\hat L_{0},\hat L_j)}{h}-\sum_{j=1}^\infty\pi_j\hat L_j\right\|\cr
&=&\left\|\frac{\sum_{i=1}^h\sum_{j=1}^\infty (\hat {L}_j\mathbb P^i(\hat L_{0},\hat L_j)-\pi_j\hat{L}_j)}{h}\right\|\cr
&=&\left\|\frac{\sum_{i=1}^h\sum_{j=1}^\infty\hat {L}_j(\mathbb P^i(\hat L_{0},\hat L_j)-\pi_j)}{h}\right\|\cr
&\le&\sup_j\|\hat {L}_j\|\frac{\sum_{i=1}^hRr^{-i}}{h}\to0,\ h\to\infty.\cr
\ean
Furthermore, by the definition of uniform convergence,   we know that
$$ \frac{1}{h}\left[\sum_{i=mh}^{mh+h-1}E[\hat{L}_{\mathcal G(i)}|\hat L_{\mathcal G(mh-1)}]\right] \mbox{ converges to} \sum_{j=1}^\infty\pi_j\hat L_j\ \mbox{a.s.}, $$
uniformly with respect to $m$, as $h\to\infty$.
Denote $\a=\lambda_2(\sum_{j=1}^\infty\pi_j\hat L_j)$. By condition (ii), it follows that $\a>0$.
Since the function $\lambda_2(\cdot)$, whose arguments are matrices, is continuous, we know that for the given $\frac{\a}{2}$, there exists a constant $\d>0$ such that for any given Laplacian matrix $L$,  $|\lambda_2(L)-\lambda_2(\sum_{j=1}^\infty\pi_j\hat L_j)|\le\frac{\a}{2}$, provided
$\|L-\sum_{j=1}^\infty\pi_j\hat L_j\|\le\d$.
Since the convergence is uniform, there exists a positive integer $h_0$ such that
$$
\left\| \frac{1}{h}\left[\sum_{i=mh}^{mh+h-1}E[\hat{L}_{\mathcal G(i)}|\hat L_{\mathcal G(mh-1)}]\right] -\sum_{j=1}^\infty\pi_j\hat L_j\right\|\le\d,\ h\geq h_0,\ \mbox{a.s.},
$$
which leads to
$$
\left|\lambda_2\left( \frac{1}{h}\left[\sum_{i=mh}^{mh+h-1}E[\hat{L}_{\mathcal G(i)}|\hat L_{\mathcal G(mh-1)}]\right]\right)-\lambda_2\left(\sum_{j=1}^\infty\pi_j\hat L_j\right)\right|\le\frac{\a}{2},\ h\geq h_0,\ \mbox{a.s.}
$$
Thus,$$\lambda_2\left( \frac{1}{h}\left[\sum_{i=mh}^{mh+h-1}E[\hat{L}_{\mathcal G(i)}|\hat L_{\mathcal G(mh-1)}]\right]\right)\ge\frac{\a}{2}>0,\ \mbox{a.s.}$$
Then by (\ref{askfjak}), we have
$\lambda_{mh}^{h}\ge\frac{h\a}{2}>0$, $h\ge h_0$ a.s.
So condition (b.1) of Theorem \ref{mainresults} holds. Then by condition (iii), we know that condition (b.2) of Theorem \ref{mainresults} holds.   Finally,  by Theorem \ref{mainresults} , we get the conclusion of the theorem. \qed

\subsection{Independent graph flow}
Consider the independent graph flow
\ban&&\Gamma_3=\Big\{\{\mathcal G(k),k\ge0\}|\{\mathcal G(k),k\ge0\}\ \mbox{is an independent process}, E[\mathcal A_{\mathcal G(k)}]\succeq O_{N\times N}, \ \mbox{a.s.} \cr&&~~~~~~~~~~~~~~~~~~~~~~~~~~~\mbox{and the associated digraph of }E[\mathcal A_{\mathcal G(k)}]\ \mbox{is balanced a.s.},\ k\geq0\Big\}.
\ean

We have the following theorem.

\begin{theorem}
\label{theoremindependentadd1}
For systems (\ref{discretemodel})-(\ref{meaequa}) and the associated random graph flow $\{{\mathcal G}(k), k\geq0\}\in\Gamma_3$, assume that

(i) Assumptions \textbf{(A1)}-\textbf{(A5)} hold;%，并且图序列$\{\mathcal

(ii) there exists a positive integer $h$ such that $$\inf_{m\ge0}\left\{\lambda_2\left[\sum_{i=mh}^{(m+1)h-1}E[\hat{L}_{\mathcal G(i)}]\right]\right\}>0;$$

(iii) $\sup_{k\ge0}E\left[\|L_{\mathcal G(k)}\|^2\right]<\infty$.

Then systems (\ref{discretemodel}))-(\ref{meaequa}) achieve mean square and almost sure average consensus.
\end{theorem}

\noindent
\textbf{Proof}: From ${\mathcal G}(k)\in\Gamma_3$, we know that  ${\mathcal G}(k)\in\Gamma_1$, and $E[\hat{L}_{\mathcal G(k)}]$ is  positive semi-definite. By the independence of $\{\mathcal G(k),k\ge0\}$, we have
$$
E[\mathcal A_{\mathcal G(k)}|\mathcal F_{\mathcal A}(k-1)]=E[\mathcal A_{\mathcal G(k)}], E[L_{\mathcal G(k)}|\mathcal F_{\mathcal A}(k-1)]=E[L_{\mathcal G(k)}],
$$
which together with Assumption \textbf{(A5)} gives
\ban
E\left[\d^T(k)\frac{ L_{\mathcal G(k)}^TP_N^T+P_N L_{\mathcal G(k)}}{2}\d(k)\right]
&=&E\left[\d^T(k)E\Big[\frac{ L_{\mathcal G(k)}^TP_N^T+P_N L_{\mathcal G(k)}}{2}|\mathcal{F}_{\xi, \mathcal{A}}(k-1)\Big]\d(k)\right]\cr
&=&E\left[\d^T(k)\frac{ E[ L_{\mathcal G(k)}^T]+ E[L_{\mathcal G(k)}]}{2}\d(k)\right]\cr
&=&E\left[\d^T(k)E[\hat{L}_{\mathcal G(k)}]\d(k)\right]\ge0.
\ean
Then similar to the proof of \textbf{Step} 1 of Theorem \ref{mainresults}, we get that $E[V(k)]$ is bounded.
Denote $\sup_{k\ge0}\left[E[\|L_{\mathcal G(k)}\|^2]\right]^{\frac{1}{2}}$ by $\rho_4$. Since $L_{\mathcal G(i)}$ is independent of $L_{\mathcal G(j)}$, $i\not=j$, we do not have to use the conditional H\"{o}lder inequality as in (\ref{random31}). Here, by the conditional Lyapunov inequality and condition (iii), we have
$\sup_{k\ge0}E[\|L_{\mathcal G(k)}\|]\le\sup_{k\ge0}\{E[\|L_{\mathcal G(k)}\|^2]\}^{\frac{1}{2}}\le\rho_{4}$.
%$$
%\sup_{k\ge0}E[\|L_{\mathcal G(k)}\|^2]\le\rho_{4}^2.
%$$
Then similar to (\ref{random31}), we obtain
\begin{eqnarray*}\label{random3100}
&&E\Big\{\Big\|\Phi^T((m+1)h,mh)\Phi((m+1)h,mh)
-I_N+\sum_{i=mh}^{(m+1)h-1}c(i)(P_NL_{\mathcal G(i)}+L_{\mathcal G(i)}^TP_N^T)
\Big\|\Big\}\cr
&\le &\Big(C_{1}\sum_{i=2}^{2h}M_{2h}^i\rho_{4}^i\Big)c^2((m+1)h)\cr
&=&C_{1}[(1+\rho_{4})^{2h}-1-2h\rho_{4}]c^2((m+1)h).
%&&\le C_{1}\rho\sum_{l=2}^{2h}M_{2h}^l c^2((m+1)h),a.s.
\end{eqnarray*}
Also by the independence of $\{\mathcal G(k),k\ge0\}$ and condition (ii), similarly to (\ref{random4}), we have
\ban\label{random5}
&&E\left[\delta^T(mh)\sum_{i=mh}^{(m+1)h-1}c(i)\left[P_NL_{\mathcal G(i)}+L_{\mathcal G(i)}^TP_N^T\right]\delta(mh)\right]\nonumber\\
&=&2E\left[\delta^T(mh)\left(\sum_{i=mh}^{(m+1)h-1}c(i)E[ \hat{L}_{\mathcal G(i)}]\right)\delta(mh)\right]\nonumber\\
&\ge&2c((m+1)h)\inf_{m\ge0}\left\{\lambda_2\left[\sum_{i=mh}^{(m+1)h-1}E[\hat{L}_{\mathcal G(i)}]\right]\right\}E[V(mh)].
%&&\ge2\theta c(k+2h) E[V(k)],
\ean
Then similarly to the proof of \textbf{Step} 2 of Theorem \ref{mainresults},  we get $E[V(k)]\to0,k\to\infty$.

By the independence of $\{\mathcal G(k),k\ge0\}$ and Assumption \textbf{(A5)}, we know that the adaptive sequences $\{\sum_{j=0}^{n}c(k)D_{\mathcal G(k)}Y(k)\xi(k)$, $\mathcal F_{\xi,{\mathcal A}}(n),n\ge0\}$ and $\{\frac{1}{N}\textbf{1}^T\sum_{k=0}^{n}c(k)
\tilde L_{\mathcal G(k)}X(k)$, $\mathcal F_{\xi,{\mathcal A}}(n),\ n\ge0\}$ are both martingale sequences.
Then similar to \textbf{Steps} 3, 4 and 5 of  Theorem \ref{mainresults}, we get the conclusion of the theorem. \qed

\vskip 0.2cm
\begin{remark}{\rm
In Theorem \ref{theoremindependentadd1}, the associated digraph of $E[\mathcal A_{\mathcal G(k)}]$, i.e., the mean graph at each time instant, is balanced, so the symmetrized mean graph is undirected. Condition (ii) of Theorem \ref{theoremindependentadd1} means that the symmetrized mean graphs are jointly-connected (the mean graph has a spanning tree) over consecutive fixed-length time intervals and the average algebraic connectivity
is uniformly positive bounded away from zero.}
\end{remark}

\vskip 0.2cm
The gossip algorithm (\cite{051}) is a special distributed averaging algorithm with an i.i.d network graph flow. For distributed averaging algorithms with random measurement noises and i.i.d graph flows, the mean square steady-state error can be estimated more precisely with sufficiently small initial algorithm gains. Moreover, the almost sure convergence rate of the $n$-step mean consensus error can be estimated.

Consider the i.i.d graph flow
\ban&&\Gamma_4=\Big\{\{\mathcal G(k),k\ge0\}|\{\mathcal G(k),k\ge0\}\ \mbox{is an i.i.d process, and} E[\mathcal A_{\mathcal G(0)}]\succeq O_{N\times N}, \mbox{and} \cr&&~~~~~~~~~~~~~~~~~~~~~~~~~~~~\mbox{ the associated digraph of }E[\mathcal A_{\mathcal G(0)}]\ \mbox{is balanced}\Big\}.
\ean

\begin{theorem}
\label{theoremiidadd1}
For systems (\ref{discretemodel})-(\ref{meaequa}) and the associated random graph flow ${\mathcal G}(k)\in\Gamma_4$, assume that

(i) Assumptions \textbf{(A1)}-\textbf{(A5)} hold;%，并且图序列$\{\mathcal

(ii) the associated digraph of the Laplacian matrix $E[L_{\mathcal G(0)}]$ has a spanning tree;

(iii) $E\left[\|L_{\mathcal G(0)}\|^2\right]<\infty$.

Then all states $x_i(k)$, $i\in\mathcal V$,  converge to a common random variable $x ^*$, in mean square and almost surely, with
$E(x^*)=\frac{1}{N}\sum_{j=1}^{N}x_j(0)$ and
\ban\label{013add}
Var(x^*)\le\frac{4c\b b^2\bar{\rho}_{1}}{N^2}+\frac{8\widetilde{c} \b\sigma^2\bar{\rho}_{1}}{N^2}+\frac{2c\bar{\rho}_2 q_x}{N},
\ean
where  $b$, $\sigma$, $c$, $\widetilde{c}$, $q_x$ are constants defined in (\ref{013}) and
\ban
\bar{\rho}_{1}=E\left[|\mathcal E_{\mathcal G(0)}|\max_{1\le i,j\le N}a_{ij}^2(0)\right],\ \bar{\rho}_2=\max_{1\le i\le N}E\left[\left(\sum_{j=1}^Na_{ij}(0)-\sum_{j=1}^Na_{ji}(0)\right)^2\right].
\ean
The convergence rate of $n$-step mean consensus error is given by
\bna
\label{meanrateaddadd1}
\frac{1}{n}\sum_{k=0}^{n}\|\delta(k)\|=o\left(\frac{1}{\sqrt{c(n)n}}\right)\ {\rm a.s.}
\ena
Furthermore, if the initial algorithm gain is so small that
\bna\label{smallcaddadd1}
c(0)<\frac{2\lambda_2\left(E[\hat{L}_{\mathcal{G}(0)}]\right)}{E\left[\|L_{\mathcal{G}(0)}\|^2\right]+4\sigma^2\beta\bar{\rho}_1},
\ena
then
\bna
\label{newboundvaraddadd1}
\widetilde{c}\leq\frac{c(0)E[V(0)]+2b^2\beta\bar{\rho}_1\sum_{k=0}^{\infty}c^3(k)}{2\lambda_2\left(E[\hat{L}_{\mathcal{G}(0)}]\right)-\left(E\left[\|L_{\mathcal{G}(0)}\|^2\right]+4\sigma^2\beta\bar{\rho}_1\right)c(0)}.
\ena
\end{theorem}

\noindent
\textbf{Proof}: It is obvious that $\Gamma_4\subseteq\Gamma_3$, so ${\mathcal G}(k)\in\Gamma_3$. By condition (ii) and ${\mathcal G}(k)\in\Gamma_4$, we know that $\lambda_2\left(E[\hat{L}_{\mathcal{G}(0)}]\right)>0$ and condition (ii) of Theorem  \ref{theoremindependentadd1} holds with $h=1$. Obviously, condition (iii) together with ${\mathcal G}(k)\in\Gamma_4$ implies condition (iii) of Theorem \ref{theoremindependentadd1}. Then by Theorem \ref{theoremindependentadd1}, the closed-loop system achieves mean square and almost sure average consensus. From (\ref{dgks}), we have
\bna
\label{nonmaraddaddadd1}
E[V(k+1)|\mathcal{F}_{\xi,\mathcal{A}}(k)]&\leq&V(k)-2c(k)\lambda_2(E[\hat{L}_{\mathcal{G}(0)}])V(k)+E[\|L_{\mathcal{G}(0)}\|^2]c^2(k)V(k)\cr
&&+4\sigma^2\beta\bar{\rho}_1c^2(k)V(k)+2b^2\beta\bar{\rho}_1c^2(k)\ {\rm a.s.},
\ena
which together with $\lambda_2\left(E[\hat{L}_{\mathcal{G}(0)}]\right)>0$ and Lemma \ref{robbins} leads to
\bna
\sum_{k=0}^{\infty}c(k)V(k)<\infty\ {\rm a.s.}
\ena
Then by Assumption \textbf{(A4)} and Kronecker lemma (\cite{044}), we know that
\ban
\lim_{n\to\infty}c(n)\sum_{k=0}^{n}V(k)=0\ {\rm a.s.},
\ean
which together with Cauchy inequality results in (\ref{meanrateaddadd1}).

From (\ref{nonmaraddaddadd1}), we have
\ban
E[V(k+1)]&\leq&E[V(k)]-2c(k)\lambda_2(E[\hat{L}_{\mathcal{G}(0)}])E[V(k)]+E[\|L_{\mathcal{G}(0)}\|^2]c^2(k)E[V(k)]\cr
&&+4\sigma^2\beta\bar{\rho}_1c^2(k)E[V(k)]+2b^2\beta\bar{\rho}_1c^2(k).
\ean
Then by Assumption \textbf{(A4)}, we have
\ban
&&(2\lambda_2(E[\hat{L}_{\mathcal{G}(0)}])-E[\|L_{\mathcal{G}(0)}\|^2]c(0)-4\sigma^2\beta\bar{\rho}_1 c(0))c^2(k)E[V(k)]\cr
&\leq&c(k)E[V(k)]-c(k+1)E[V(k+1)]+2b^2\beta\bar{\rho}_1c^3(k).
\ean
Take summation on both sides of the above inequality from $k=0$ to $k=n$ gives
\ban
&&(2\lambda_2(E[\hat{L}_{\mathcal{G}(0)}])-E[\|L_{\mathcal{G}(0)}\|^2]c(0)-4\sigma^2\beta\bar{\rho}_1 c(0))\sum_{k=0}^{n}c^2(k)E[V(k)]\cr
&\leq&c(0)E[V(0)]-c(n+1)E[V(n+1)]+2b^2\beta\bar{\rho}_1\sum_{k=0}^{n}c^3(k).
\ean
Then by (\ref{smallcaddadd1}) and let $n\to\infty$, we have (\ref{newboundvaraddadd1}). \qed

\section{conclusion}
We have considered discrete-time stochastic approximation type distributed averaging algorithms with random measurement noises and time-varying random graph flows. Compared with the existing literature, our model is more widely applicable in the sense that i) the measurement covers both additive and multiplicative noises; ii) the network graphs and noises are not required to be spatially and temporally independent; iii) the edge weights of network graphs are not necessarily nonnegative with probability one. By constructing difference inequalities of proper stochastic Lyapunov function, the algebraic graph theory and martingale convergence theory,  sufficient conditions have been given to achieve mean square and almost sure average consensus. It has been shown that all states of agents converge to a common variable in mean square and almost surely if the network graph flow is \emph{conditionally balanced and uniformly conditionally jointly connected}. The mathematical expectation of the common variable is right the average of initial values. Moreover, an upper bound of the mean square steady-state error has been given in relation to the edge weights, the time-varying algorithm gain, the number of agents, the agents' initial values, the second-order moment and the intensity coefficients of noises. Especially, if the measurement noises are both spatially and temporally independent, then the mean square steady-state error vanishes as the number of nodes increases to infinity under mild conditions on the network graphs.

Convergence rate is an important performance for distributed averaging algorithms. Different from the fixed-gain algorithms for noise-free cases  (\cite{BXMS2013TSP}, \cite{PBA2010TAC}, \cite{OT2009SIAM}), here,  the non-zero off-diagonal elements  of the closed-loop state matrix are not uniformly bounded away from zero, which results in much more difficulties to get the exact stochastic convergence rates of the algorithm. For the case with i.i.d graph flows, we have given a rough estimate for the $n$-step mean consensus error with probability one. It is interesting to develop effective tools to give the exact stochastic convergence rates of our algorithms.

%\noindent
\section*{Appendix}
\setcounter{lemma}{0}
\def\thelemma{A.\arabic{lemma}}
\setcounter{definition}{0}
\def\thedefinition{A.\arabic{definition}}
\setcounter{equation}{0}
\def\theequation{A.\arabic{equation}}
\begin{lemma}\label{independence}
Let $\{Z_k,k\ge0\}$ and $\{W_k,k\ge0\}$ be mutually independent random vector sequences. Then $\sigma(Z_j,Z_{j+1},...)$ and $\sigma(W_j,W_{j+1},...)$ are conditionally independent given $\sigma(Z_0,...,Z_{j-1},W_0,...,W_{j-1})$,  $\forall\ j\geq1$.%$\sigma_1=\sima(Z_j,j\ge k)$和$\sigma_2=\sigma(W_j,j\ge k)$
\end{lemma}

\noindent
\textbf{Proof}: Denote $Z_{m\sim n}=\{Z_m=z_m,...,Z_{n}=z_{n}\}$ and $Z_{m\sim \infty}=\{Z_m=z_m, Z_{m+1}=z_{m+1},...\}$ where $z_k$ denotes the possible values of $Z_k$.
By the definition of conditional probability,
\bna\label{conditionalproof2}
&&\mathbb P\{Z_{j\sim\infty}, W_{j\sim\infty}|Z_{0\sim j-1},W_{0\sim j-1}\}\cr
&=&\mathbb P\{W_{j\sim\infty}|Z_{0\sim j-1},W_{0\sim j-1}\} \mathbb P\{Z_{j\sim\infty}|Z_{0\sim j-1},W_{0\sim\infty}\}.
\ena
Noting that $\sigma(Z_{0\sim\infty})=\sigma(\sigma(Z_{j\sim\infty})\cup\sigma(Z_{0\sim j-1}))$ and $\sigma(Z_{0\sim\infty})$ is independent of $\sigma(W_{0\sim\infty})$, by Corollary 3 of Section 7.3 of \cite{044}, we have
\ban
\mathbb P\{Z_{j\sim\infty}|Z_{0\sim j-1},W_{0\sim\infty}\}=
\mathbb P\{Z_{j\sim\infty}|Z_{0\sim j-1}\}=\mathbb P\{Z_{j\sim\infty}|Z_{0\sim j-1},W_{0\sim j-1}\},
\ean
which together with (\ref{conditionalproof2}) gives
\ban\label{conditionalproof1}
\mathbb P\{Z_{j\sim\infty},W_{j\sim\infty}|Z_{0\sim j-1},W_{0\sim j-1}\}
=\mathbb P\{W_{j\sim\infty}|Z_{0\sim j-1},W_{0\sim j-1}\}
\mathbb P\{Z_{j\sim\infty}|Z_{0\sim j-1},W_{0\sim j-1}\}.
\ean
By the definition of conditional independence, we get the conclusion. \qed

\vskip 0.2cm

\begin{lemma}(\cite{rb})\label{robbins}
Let $\{x(k), \mathcal F(k)\}$, $\{\a(k),\mathcal F(k)\}$, $\{\b(k),\mathcal F(k)\}$ and $\{\gamma(k), \mathcal F(k)\}$ be nonnegative adaptive sequences  satisfying
$$
E(x(k+1)|\mathcal F(k))\le(1+\a(k))x(k)-\beta(k)+\gamma(k),\ k\ge0\ a.s.,
$$
and $\sum_{k=0}^\infty(\a(k)+\gamma(k))<\infty$ a.s.  then $x(k)$ converges to a finite random variable a.s., and $\sum_{k=0}^\infty\b(k)<\infty$ a.s.
\end{lemma}

\vskip 0.2cm

\begin{lemma}\label{lemma1}(\cite{033})
Let $\{u(k)$, $k\ge0\}$, $\{q(k)$, $k\ge0\}$ and $\{\a(k)$, $k\ge0\}$ be real sequences, where $0<q(k)\le 1$, $\a(k)\ge0$, $k\ge0$, $\sum_{k=0}^\infty q(k)=\infty$, $\frac{\a(k)}{q(k)}\to0$, $k\to\infty$,
and $u({k+1})\le(1-q(k))u(k)+\a(k)$. Then $\limsup_{k\to\infty}u(k)\le0$. Especially, if $u(k)\ge0$, $k\ge0$, then $u(k)\to0$, $k\to\infty$.
\end{lemma}

\vskip 0.2cm

\begin{lemma}\label{aksfmo}(\cite{046})
Let $\{X(k),\mathcal F(k)\}$ be a martingale sequence satisfying $\sup_{k\ge0}E[\|X(k)\|^2]<\infty$. Then $X(k)$ converges in mean square and almost surely.
\end{lemma}

%\end{CJK*}
\end{document}